\documentclass[longauth]{aa}
\usepackage[varg]{txfonts}
 
\usepackage{amsmath}
\usepackage{amsfonts}
\usepackage{amssymb}
\usepackage{graphicx}
\usepackage{fourier}

\usepackage{natbib}
\bibpunct{(}{)}{;}{a}{}{,} % to follow the A&A style

\usepackage{booktabs} % For the tables

\usepackage{makecell}
\usepackage{txfonts}
\usepackage{graphicx,natbib,url,twoopt}
\usepackage{lipsum}
\usepackage{multicol}
\usepackage{float}

\usepackage{stackengine}

\usepackage{threeparttable}

\usepackage{subfig}

\usepackage{longtable}

\usepackage{placeins}
\def\logg{\hbox{$\log{g}$}}         %logg
\def\rhk{\hbox{$\log{(R^{'}_{HK})}$}}         %rhk
\def\ha{\hbox{$H_{\alpha}$}}         %Halpha
\def\m2s2{\hbox{\,m$^{2}$\,s$^{-2}$}} %m2.s -2
\def\teff{\hbox{$T_{\rm eff}$}}

\begin{document}

\title{TESS and HARPS-N unveil two planets \\ transiting TOI-1453}
\subtitle{A super-Earth and one of the lowest mass sub-Neptunes\thanks{The results presented in this paper have been obtained within the HARPS-N Collaboration.}}

\author{M. Stalport\inst{\ref{i:liege_STAR}, \ref{i:liege_EXOTIC}}
\and A. Mortier\inst{\ref{i:Birmingham}}
\and M. Cretignier\inst{\ref{i:Oxford}}
\and J. A. Egger\inst{\ref{i:Bern}} 
\and L. Malavolta\inst{\ref{i:Padova}, \ref{i:INAF-Padova}}
\and D. W. Latham\inst{\ref{i:CfA}}
\and K. A.\ Collins \inst{\ref{i:CfA}} 
\and C. N.\ Watkins \inst{\ref{i:CfA}}
\and F. Murgas\inst{\ref{i:IAC-Tenerife}, \ref{i:ULL-Tenerife}} 
\and L. A. Buchhave\inst{\ref{i:Lyngby}}
\and M. López-Morales\inst{\ref{i:CfA}}
\and S. Udry\inst{\ref{i:Geneva}}
\and S. N. Quinn\inst{\ref{i:MIT}}
\and A. M. Silva\inst{\ref{i:Porto_IACU}, \ref{i:Porto}}
\and G. Andreuzzi\inst{\ref{i:INAF-FGG}, \ref{i:INAF-Rome}}
\and D. Baker\inst{\ref{i:AustinColl}}
\and W. Boschin\inst{\ref{i:INAF-FGG}, \ref{i:IAC-Tenerife}, \ref{i:ULL-Tenerife}}
\and D. R. Ciardi\inst{\ref{i:Caltech-IPAC}}
\and M. Damasso\inst{\ref{i:INAF-Torino}}
\and L. Di Fabrizio\inst{\ref{i:INAF-FGG}}
\and X. Dumusque\inst{\ref{i:Geneva}}
\and A. Fukui\inst{\ref{i:Tokyo}}
\and R. Haywood\inst{\ref{i:Exeter}}
\and S. B. Howell\inst{\ref{i:NASA}}
\and J. M. Jenkins\inst{\ref{i:NASA}}
\and A. Leleu\inst{\ref{i:Geneva}}
\and P. Lewin\inst{\ref{i:MauryLewinObs}}
\and J. Lillo-Box\inst{\ref{i:CAB-Madrid}}
\and A. F. Martínez Fiorenzano\inst{\ref{i:INAF-FGG}}
\and N. Narita\inst{\ref{i:Tokyo}} 
\and M. Pedani\inst{\ref{i:INAF-FGG}}
\and M. Pinamonti\inst{\ref{i:INAF-Torino}}
\and E. Poretti\inst{\ref{i:INAF-FGG}}
\and R. P. Schwarz\inst{\ref{i:CfA}}
\and S. Seager\inst{\ref{i:MIT}, \ref{i:MIT_EAPS}, \ref{i:MIT_AA}}
\and A. Sozzetti \inst{\ref{i:INAF-Torino}} 
\and E. B. Ting\inst{\ref{i:NASA}}
\and A. Vanderburg\inst{\ref{i:MIT_PHYS}}
\and J. N.\ Winn\inst{\ref{i:Princeton}} 
\and C. Ziegler\inst{\ref{i:ASU}}
}

\institute{Space sciences, Technologies and Astrophysics Research (STAR) Institute, Université de Liège, Allée du 6 Août 19C, 4000 Liège, Belgium \label{i:liege_STAR} 
\and Astrobiology Research Unit, Universit\'e de Li\`ege, All\'ee du 6 Août
19C, B-4000 Liège, Belgium \label{i:liege_EXOTIC}
\and School of Physics \& Astronomy, University of Birmingham, Edgbaston, Birmingham B15 2TT, UK \label{i:Birmingham} 
\and Department of Physics, University of Oxford, OX13RH Oxford, UK \label{i:Oxford}
\and Weltraumforschung und Planetologie, Physikalisches Institut, University of Bern, Gesellschaftsstrasse 6, 3012 Bern, Switzerland \label{i:Bern} 
\and Dipartimento di Fisica e Astronomia ‘Galileo Galilei’, Universitá di Padova, Vicolo del l’Osservatorio 3, 35122 Padova, Italy \label{i:Padova}
\and INAF -- Osservatorio Astronomico di Padova, Vicolo dell’Osservatorio 5, Padova, 35122 Italy \label{i:INAF-Padova}
\and Center for Astrophysics \textbar{} Harvard \& Smithsonian, 60 Garden Street, Cambridge, MA 02138, USA \label{i:CfA}
\and Instituto de Astrof\'{\i}sica de Canarias, C/V\'{\i}a L\'actea s/n, E-38205 La Laguna (Tenerife), Canary Islands, Spain \label{i:IAC-Tenerife}
\and Departamento de Astrof\'{\i}sica, Univ. de La Laguna, Av. del Astrof\'{\i}sico Francisco S\'anchez s/n, E-38205 La Laguna (Tenerife), Canary Islands, Spain \label{i:ULL-Tenerife}
\and DTU Space, Technical University of Denmark, Elektrovej 328, DK-2800 Kgs. Lyngby, Denmark \label{i:Lyngby}
\and Département d’Astronomie, Université de Genève, Chemin Pegasi 51b, 1290 Versoix, Switzerland \label{i:Geneva} 
\and Department of Physics and Kavli Institute for Astrophysics and Space Research, Massachusetts Institute of Technology, Cambridge, MA 02139, USA \label{i:MIT}
\and Instituto de Astrofísica e Ciências do Espaço, Universidade do
Porto, CAUP, Rua das Estrelas, 4150-762 Porto, Portugal \label{i:Porto_IACU}
\and Departamento de Física e Astronomia, Faculdade de Ciências,
Universidade do Porto, Rua do Campo Alegre, 4169-007 Porto,
Portugal \label{i:Porto}
\and Fundaci\'on Galileo Galilei -- INAF (Telescopio Nazionale Galileo), Rambla Jos\'e Ana Fern\'andez Perez 7, 38712 Breña Baja (La Palma), Canary Islands, Spain \label{i:INAF-FGG}
\and INAF -- Osservatorio Astronomico di Roma, Via Frascati 33, 00078 Monte Porzio Catone, Italy \label{i:INAF-Rome}
\and Physics Department, Austin College, Sherman, TX 75090, USA \label{i:AustinColl} 
\and NASA Exoplanet Science Institute, IPAC, California Institute of
Technology, Pasadena, CA 91125 USA \label{i:Caltech-IPAC}
\and INAF -- Osservatorio Astrofisico di Torino, Via Osservatorio 20, 10025 Pino Torinese, Italy \label{i:INAF-Torino}
\and Komaba Institute for Science, The University of Tokyo, 3-8-1 Komaba,
Meguro, Tokyo 153-8902, Japan \label{i:Tokyo}
\and Astrophysics Group, University of Exeter, Exeter EX4 2QL, UK \label{i:Exeter}
\and NASA Ames Research Center, Moffett Field, CA 94035, USA \label{i:NASA} 
\and The Maury Lewin Astronomical Observatory, Glendora, CA 91741, USA \label{i:MauryLewinObs} 
\and Centro de Astrobiología (CAB), CSIC-INTA, ESAC campus, Camino Bajo del Castillo s/n, 28692 Villanueva de la Cañada (Madrid), Spain \label{i:CAB-Madrid} 
\and Department of Earth, Atmospheric, and Planetary Sciences, Massachusetts Institute of Technology, Cambridge, MA 02139, USA \label{i:MIT_EAPS} 
\and Department of Aeronautics and Astronautics, Massachusetts Institute of Technology, Cambridge, MA 02139, USA \label{i:MIT_AA} 
\and Department of Physics and Kavli Institute for Astrophysics and Space Research, Massachusetts Institute of Technology, Cambridge, MA 02139, USA \label{i:MIT_PHYS}
\and Department of Astrophysical Sciences, Princeton University, Princeton, NJ 08544, USA \label{i:Princeton}
\and Department of Physics, Engineering and Astronomy, Stephen F. Austin State University, 1936 North St, Nacogdoches, TX 75962, USA \label{i:ASU}
}

\date{Received 12 November 2024 / Accepted 23 February 2025}

% **************************************************** ABSTRACT 
\abstract
{The TESS mission is searching for transiting planets over the entire sky, including two continuous viewing zones. Data from the continuous viewing zones span a long time baseline and offer ideal conditions for precise planet radius estimations, enabling the community to prepare for the PLATO mission.}
{We report on the validation and characterisation of two transiting planets around TOI-1453, a K-dwarf star in the TESS northern continuous viewing zone.}
{In addition to the TESS data, we used ground-based photometric, spectroscopic, and high-resolution imaging follow-up observations to validate the two planets. We obtained 100 HARPS-N high-resolution spectra over two seasons and used them together with the TESS light curve to constrain the mass, radius, and orbit of each planet.}
{TOI-1453\,b is a super-Earth with an orbital period of $P_b$=4.314 days, a radius of $R_b$=1.17$\pm$0.06\,$R_{\oplus}$, and a mass lower than 2.32\,$M_{\oplus}$ (99$\%$). 
TOI-1453\,c is a sub-Neptune with a period of $P_c$=6.589 days, radius of $R_c$=2.22$\pm$0.09\,$R_{\oplus}$, and mass of $M_c$=2.95$\substack{+0.83 \\ -0.84}$\,$M_{\oplus}$. The two planets orbit TOI-1453 with a period ratio close to 3/2, although they are not in a mean motion resonance (MMR) state. We did not detect any transit timing variations in our attempt to further constrain the planet masses. TOI-1453\,c has a very low bulk density and is one of the least massive sub-Neptunes discovered to date. It is compatible with having either a water-rich composition or a rocky core surrounded by a thick H/He atmosphere. However, we set constraints on the water mass fraction in the envelope according to either a water-rich or water-poor formation scenario. The star TOI-1453 belongs to the Galactic thin disc based on \textit{Gaia} kinematics and has a sub-solar metallicity. This system is orbited by a fainter stellar companion at a projected distance of $\sim$150\,AU, classifying TOI-1453\,b and c of S-type planets. These various planetary and stellar characteristics make TOI-1453 a valuable system for understanding the origin of super-Earths and sub-Neptunes.}
{}{}


\maketitle

\nolinenumbers

% **************************************************** INTRODUCTION 
\section{Introduction} \label{Sect:Intro} 
Super-Earths and sub-Neptunes represent a class of planets that do not exist in the Solar System. Theoretical works have  sought to understand why such planets do not orbit the Sun. Notably, it has been proposed that Jupiter played a role in inhibiting the formation of super-Earths and sub-Neptunes \citep{Izidoro2015, Lambrechts2019}. With estimated radii in the range 1$R_{\oplus}$ < $R_p$ < 4$R_{\oplus}$, the discovery of these planets around other stars was the result of large surveys, coupled with important instrumental developments. Radial velocity (RV) blind searches with high-resolution spectrographs \citep[e.g. HARPS,][]{Pepe2000}, together with the space-based Kepler transit survey \citep{Borucki2010} have revealed this new category of planets that is intermediate in mass and radius to Earth and Neptune \citep{Mayor2008}. 
Since then, the number of characterised super-Earths and sub-Neptunes has increased dramatically thanks to the Transiting Exoplanet Survey Satellite \citep[TESS,][]{Ricker2015}. The efficient coordination with ground-based RV follow-up of transiting exoplanet candidates has allowed the community to fill in the mass-radius space and has provided important insights into the composition of these planets. While super-Earths and sub-Neptunes overlap in their range of masses \citep[e.g.][]{Otegi2020, Parc2024}, they show a more distinct separation in radius \citep{Fulton2017, Mayo2018, Cloutier2020}. This separation, known as the radius valley, suggests also a separation in bulk density, as proposed by \citet{Luque2022} for planets around M-dwarf stars. This separation in density has been more recently supported by pebble-based formation models for all stellar types \citep{Venturini2024}; however, the question of a structural difference between various populations of sub-Neptunes is still heavily debated \citep[e.g.][]{Rogers2023, Parc2024}. 
Understanding the composition of super-Earths and sub-Neptunes, as well as the origin for such differences, will bring important insights into the general theory of planet formation and evolution. 

Several models attempt to explain the emergence of the observed differences between super-Earths and sub-Neptunes. Photo-evaporation \citep{Owen2013} and core-powered mass loss \citep{Ginzburg2018} are two popular hypotheses positing that some planets would have lost most of their atmospheres to become the super-Earths we observe today. Other hypotheses explore in situ formation of super-Earths, where the planets would form at a later stage in a gas-poor environment \citep{Lopez2018}. These various models make different predictions regarding the location of the valley in the orbital period-radius space \citep[e.g.][]{Cloutier2020}. Therefore, a continued effort to detect and characterise super-Earths and sub-Neptunes is of utmost interest. Among the newly discovered planets, multi-planet systems are particularly valuable. They provide an opportunity to compare planet properties within the same formation environment and around the same star and to compare these characteristics to dynamical processes. Multi-planet systems are key laboratories to interpret population-level studies and constrain formation and evolution models. 

Within this observational effort, the high-resolution echelle spectrograph HARPS-N \citep{Cosentino2012} plays a key role. With a proven stability at the sub-ms$^{-1}$ level \citep{Cosentino2014, John2023}, HARPS-N has played a significant role in measuring the masses of small planet candidates identified with data from the Kepler \citep{Borucki2010}, K2 \citep{Howell2014}, and TESS \citep{Ricker2015} missions. As of the 14th of May 2024, the HARPS-N Collaboration has measured the mass of more than a third of the transiting small planets ($R_p$<4$R_{\oplus}$) with a precision better than 30$\%$, playing a major role in populating the mass-radius diagram \citep[e.g.][]{Pepe2013, Malavolta2018, Mortier2020, Cloutier2020b, Lacedelli2021, Rajpaul2021, Bonomo2023, Palethorpe2024}. 

In this paper, we report on the discovery and characterisation of a super-Earth and a sub-Neptune orbiting the K-dwarf TOI-1453 (TIC 198390247). The system is located in the TESS northern continuous viewing zone (CVZ) and benefits from a wealth of photometric data. We carried out follow-up RV observations with HARPS-N during two seasons. The combination of these two data sets allowed us to constrain the planetary masses and radii. The planets span the radius valley and have a ratio of orbital periods of nearly 3 to 2. Additionally, the system is orbited by a faint stellar companion. This system therefore contains the ingredients to set important constraints on the formation and evolution of super-Earths and sub-Neptunes. 

In Sect. \ref{Sect:Observations}, we describe the ensemble of observations that we used to both validate the planet candidates and characterise them. Section \ref{Sect:StellarParameters} presents the astrometric and atmospheric characterisation of the star, as well as an analysis of its activity. We then tackle the data analysis in Sect. \ref{Sect:DataAnalysis} so to constrain the planet physical parameters. In Sect. \ref{Sect:DynamicalState}, we study the dynamics of this two-planet system and investigate the transit timing variations (TTVs). We explore the possible structures of the sub-Neptune TOI-1453 c in Sect. \ref{Sect:Interior}. Finally, we discuss the place of TOI-1453 in the context of an exoplanet population and present the main takeaways of this peculiar planetary system in Sect. \ref{Sect:Conclusion}.

% **************************************************** OBSERVATIONS 
\section{Observations} \label{Sect:Observations} 
This section summarises the ensemble of TOI-1453 observations used in the present work. 
\subsection{TESS photometry} \label{Sect:Obs:TESS}
The TESS observations of this star started on 2019 July 18 (sector 14), and have since then covered a large number of sectors. As of April 2024, TESS has observed TOI-1453 in sectors 14-17, 19-26, 40-41, 47, 49-54, 56-57, 59-60, and 73-76, for a total of 29 sectors and a baseline of 1711 days. All the data are available with a 2 minute cadence. We note also that 20 second cadence data are available for this star on the sectors comprised between 40 and 60. These data were processed by the Science Processing Operations Center at NASA Ames Research Center \citep[SPOC; ][]{Jenkins2016}. 

An early report from the SPOC based on sectors 14, 15, and 16 identified a candidate transiting planet with an orbital period of 13.18 days. This finding was obtained using the Transiting Planet Search (TPS) pipeline module \citep{Jenkins2002, Jenkins2010, 2020TPSkdph}. However, the phase-folded light curve presented a significant flux decrease at exactly half the period, suggesting that the true orbital period could be 6.59 days. The orbital period of this candidate was updated to 6.59 days in the SPOC report for sector 21, with a corresponding transit depth of 894 $\pm$ 228 ppm. MIT’s Quick Look Pipeline \citep[QLP,][]{Huang2020a, Huang2020b} also performed a transiting planet search of sectors 14-16 using FFI data and detected the signature of TOI-1453\,c at the correct period of 6.59 days. The SPOC fitted an initial limb-darkened transit model to the signature \citep{Li2019} which passed a suite of diagnostic tests to help make or break the planetary nature of the signal \citep{Twicken2018}. The TESS Science Office (TSO) reviewed the results and issued an alert for TOI-1453\,c on 14 November 2019 \citep{Guerrero2021}. In the meantime, the SPOC Data Validation report for sectors 14-19 detected the signature of a small second planetary candidate with an orbital period of 4.31 days and a transit depth of 243 $\pm$ 243 ppm, and the TSO issued an alert for TOI-1453 b on 21 February 2020. Beyond detecting the signatures of TOI-1453\,b and c, the SPOC difference image centroiding tests conducted for the search of sectors 14-55 constrained the location of the host star to within 5.0±5.5 arcsec of the transit source of TOI-1453 b, and to within 1.4±3.4  arcsec of the transit source of TOI-1453 c, respectively.

\begin{figure} 
    \centering
    \includegraphics[width=\columnwidth]{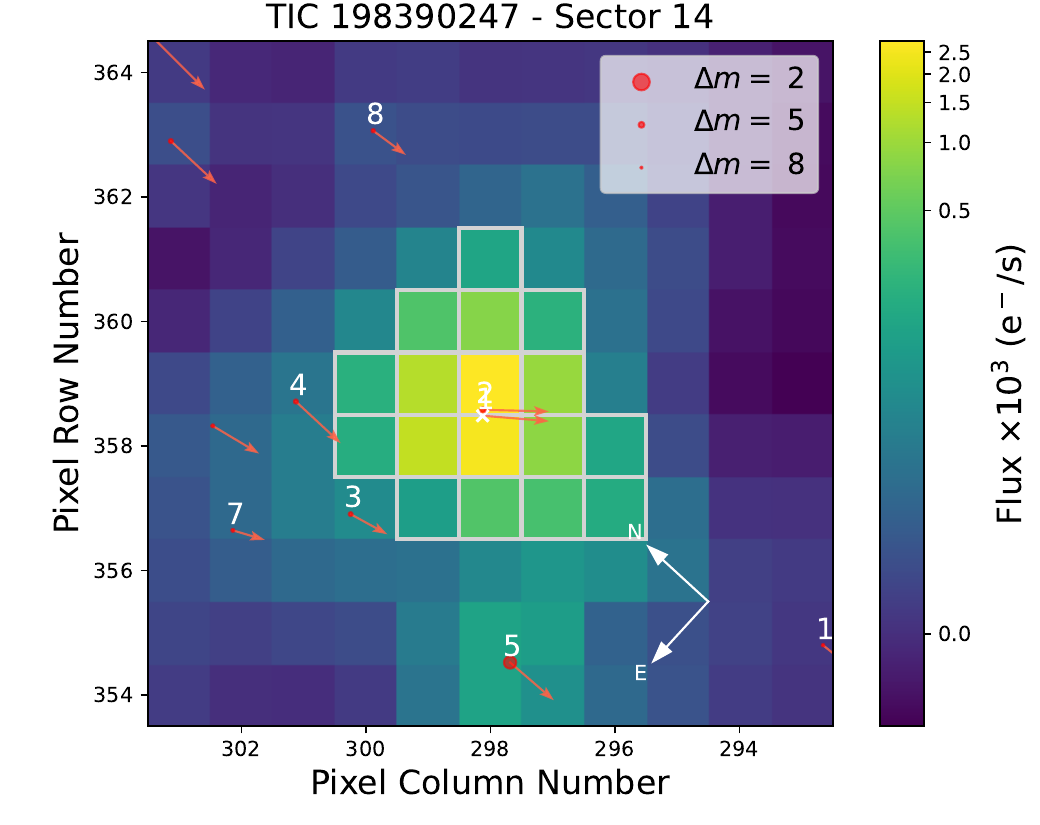}
\caption{Example of TESS target pixel file centred on TOI-1453 (TIC 198390247). The target is highlighted with a white cross at the centre of the grid. The zone inside of which we perform aperture photometry is contoured in bold gray. All the sources in the neighbourhood of TOI-1453 identified with the \textit{Gaia} DR3 catalogue are shown in red, and the arrows illustrate their proper motions on the sky. This plot was designed with \texttt{tpfplotter} \citep{Aller2020}.}
\label{Fig:TPF} 
\end{figure}

Figure \ref{Fig:TPF} shows a sample TESS target pixel file from sector 24, illustrating the low spatial resolution of TESS (the pixel scale is 21$\arcsec$). As a result, blending with other sources is expected. The pipeline photometric aperture is selected to maximise the signal-to-noise ratio (S/N) of the target star. This is highlighted in Fig.\,\ref{Fig:TPF} with light gray thick contours to the pixels. In this work, we employed the pipeline photometric aperture to retrieve the light curves. Some blended sources may be hidden by the target on the image. We therefore identified all the known nearby sources from the \textit{Gaia} DR3 catalogue and marked these close sources in red on the figure. In particular, a very close source appears inside the pipeline photometric aperture. It has a relative \textit{Gaia} magnitude of $\Delta G$=5.24 with respect to the target and a similar proper motion to TOI-1453, indicative of a gravitationally bound object. 

To retrieve the light curves, we employed a custom extraction from the pipeline aperture of the target pixel files. Using \texttt{lightkurve} \citep{Lightkurve2018}, we removed the background scattered light by applying a principal components analysis on the out-of-aperture pixels matrix and decorrelating the target flux from the five main components via a linear regression. Additionally, we removed the most common systematic trends observed in each TESS CCD and for each sector via the co-trending basis vectors (CBVs), which are generated in the pre-search data conditioning (PDC) component of the TESS pipeline. In particular, we employed a combination of multi-scale and spike CBVs, in order to correct for the systematics on different wavelet-based band passes as well as the short impulsive systematics. For each band pass of the multi-scale CBVs, we used eight CBVs. We extracted and corrected the light curve from systematics with the joint background correction and the CBV detrending, using the function \texttt{CBVCorrector.correct$\_$gaussian$\_$prior} in \texttt{lightkurve}. The resulting light curve did not need further detrending. Indeed, we performed a model comparison on each sector, between models with and without Gaussian process (GP) detrending. These analyses were carried out with the \texttt{Juliet} Python package \citep{juliet}, which employs the code \texttt{batman} \citep{batman} to fit for the two-planet transit model. We tested different GP kernels in our model comparison, each of them implemented in the code \texttt{celerite} \citep{celerite} which \texttt{Juliet} depends on. The model parameters of the fit were explored using the nested sampling algorithm \texttt{dynesty} \citep{Speagle2020}, which also delivers the Bayesian evidence of the model. Unambiguously, our analyses have provided more evidence to support use of the simpler model without GP detrending. 

We also compared the resulting light curve with the PDC simple aperture photometry \citep[PDCSAP; ][]{Stumpe2012, Smith2012, Stumpe2014}. We noticed an overall improvement with our custom extraction, with a smaller, or at worst equivalent, scatter in the individual sectors. Furthermore, because of the quality flag applied in the PDCSAP light curve, a significant number of transits are missed. With our custom extraction, we could retrieve 21 additional transits of the inner planet candidate and 10 additional transits of the outer candidate over all the sectors with a minor impact on the light curve quality, as noted with the scatter comparison. In the following, we use the light curve obtained from our custom extraction. We provide a synthesis of each sector of our dataset in Table \ref{tab:TESS_sectors}.

\subsection{High-resolution imaging}
High-resolution images were taken from the ground to search for nearby stars that might be blended with TOI-1453 in the TESS images. Their identification is important not only to identify potential false positive transits, but also to update the light dilution correction factor when needed. High-resolution images of TOI-1453 and its surrounding field were obtained with three different instruments and telescopes. 

\begin{figure} 
    \centering
    \includegraphics[width=\columnwidth]{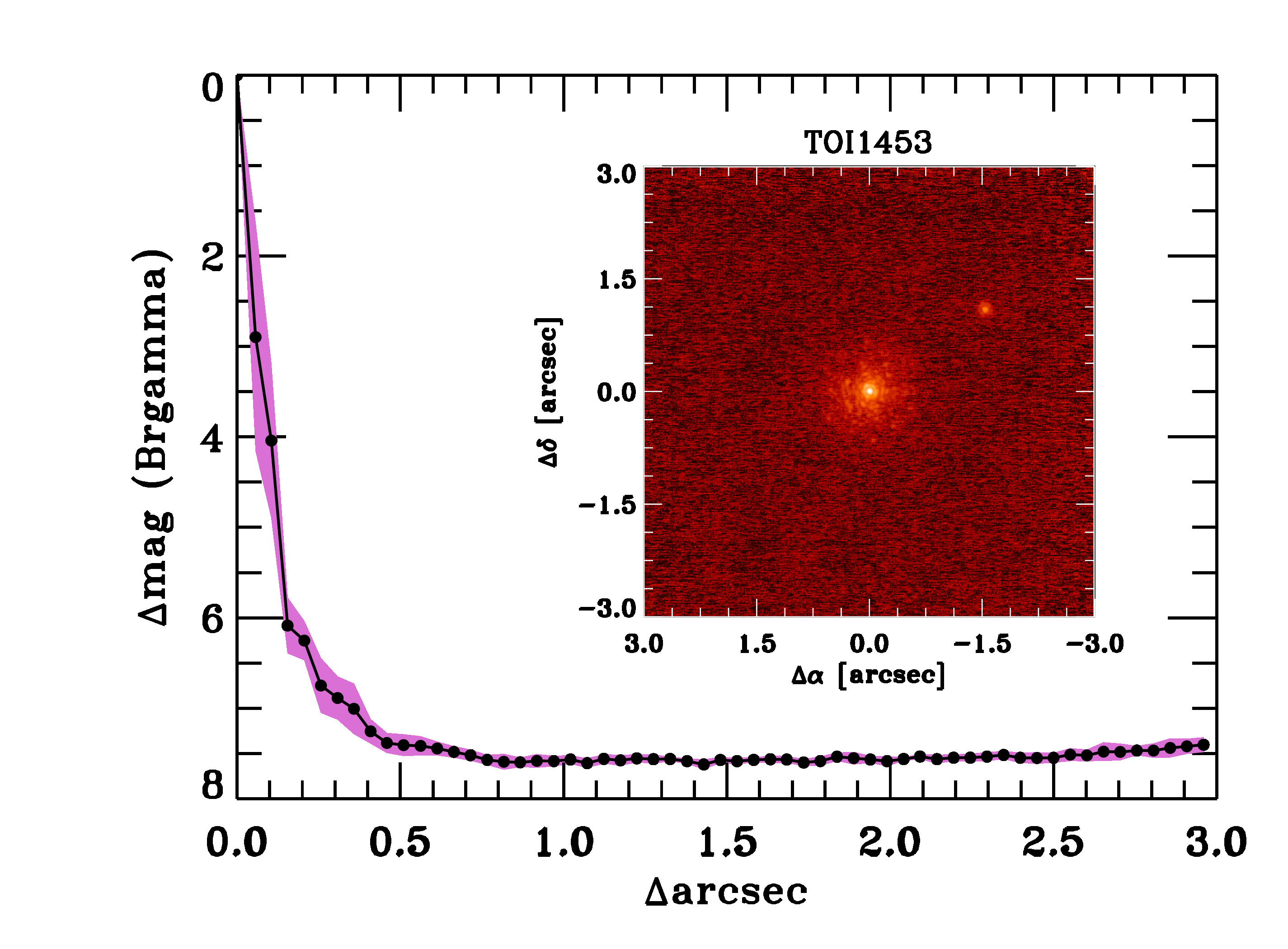}
\caption{AO image of TOI-1453 and corresponding sensitivity curve obtained with NIRC2 installed on the Keck2 telescope (cf. \url{https://exofop.ipac.caltech.edu/tess/view_tag.php?tag=19377}, PI: Gonzales). The filter used for this image is $B_r\gamma$. The faint companion at about 2\arcsec away from TOI-1453 is revealed by this image.} 
\label{Fig:AO_Keck} 
\end{figure}

Two adaptive optics (AO) images were obtained on 2020 May 28 with the NIRC2 instrument coupled with the Keck2 10m telescope located at Mauna Kea observatory, Hawaii. One of the images was taken in the $J_{cont}$ photometric band (1.21 $\mu$m), the other in the $B_r\gamma$ filter (2.17 $\mu$m). These images have a resolution of 9.942 milliarcsec/pixel, and have a contrast detection threshold at 0.5\arcsec angular separation of $\Delta$mag = 6.43 and 7.41, respectively, for the $J_{cont}$ and $B_r\gamma$ filters. Figure \ref{Fig:AO_Keck} presents the contrast curve together with the resulting image, obtained in the $B_r\gamma$ filter. TOI-1453 appears at the center of the image, and its faint stellar companion is seen about 2\arcsec away. 

Two speckle images were obtained on 2020 June 6 with the Alopeke instrument installed on the Gemini 8m telescope at Mauna Kea observatory, Hawaii. The achieved resolution of this data is 10 mas/pixel. Each image was obtained in a distinct narrow photometric band, around 562\,nm and 832\,nm, for which the contrast threshold at 0.5\arcsec is measured to $\Delta$mag = 4.83 and 6.62, respectively. 

Finally, lucky images of TOI-1453 were taken on 2021 March 22 and 2021 September 29, using the AstraLux instrument on the 2.2m telescope at Calar Alto observatory. The observations were carried out in the SDSSz bandpass, and present a resolution of 23.27 milliarcsec/pixel. The first observation taken in March 2021 has a contrast detection threshold at 1\arcsec of $\Delta$mag = 5.6, while the second observation obtained in September 2021 has a threshold contrast of 4.3. 

All these high-resolution observations jointly confirm that besides the known companion 1.9\arcsec away from TOI-1453, it is most likely that the star has no other close stellar companion. 
A mismatch was identified by \citet{LilloBox2024} between the position of the stellar companion in the AstraLux lucky images and the \textit{Gaia} data. The latter indicates a separation of 3.02\arcsec, while they report a separation of 1.85\arcsec with AstraLux (consistent with the other high-resolution images). The origin of this discrepancy is yet unclear, and might indicate that the companion star moved between the \textit{Gaia} epoch and 2020-2021.

\subsection{TRES reconnaissance spectroscopy}
Early on after the planet candidate alert released by the TSO, two spectra of TOI-1453 were obtained with the Tillinghast Reflector Echelle Spectrograph (TRES) installed on the 1.5m telescope at Fred L. Whipple Observatory (FLWO), Arizona. The two observations were performed on 2020 January 30 and 2020 March 16. The aim of those observations was to confirm the single-line nature of the spectrum, assess the activity level of the star and its convenience for precise RV follow-up, and to reject false positive hypotheses. TRES presents a spectral resolution of 44\,000 in the broad visible range (from 385 to 909.6\,nm). The first observation revealed a single-line spectrum, appropriate for precise RV follow-up. Because of the initial confusion regarding the true orbital period of TOI-1453.01, the second TRES spectrum was taken at the same phase as the first one. A negligible velocity shift was noted.

\subsection{Ground-based photometry} \label{Sect:GroundBasedPhot}
The TESS pixel scale is $\sim 21\arcsec$ pixel$^{-1}$ and photometric apertures typically extend out to roughly 1 arcminute, generally causing multiple stars to blend in the TESS photometric aperture. In our attempt to determine the true source of the TESS detection, we acquired follow-up ground-based photometric observations of the fields around TOI-1453 as part of the TESS Follow-up Observing Program \citep[TFOP;][]{Collins2019}\footnote{\url{https://tess.mit.edu/followup}}. 

\subsubsection{LCOGT}
We used the {\tt TESS Transit Finder}, which is a customised version of the {\tt Tapir} software package \citep{Jensen2013}, to schedule our transit observations. We observed two full transit windows of TOI-1453.01 (candidate planet c) in Pan-STARRS $z$-short band on UTC 2020 May 17 and 2020 August 24 from the Las Cumbres Observatory Global Telescope (LCOGT) \citep{Brown2013} 1\,m network node at McDonald Observatory near Fort Davis, Texas, United States (McD). The observation on 2020 August 24 was simultaneously observed with two 1\,m telescopes at McD. The 1\,m telescopes are equipped with a $4096\times4096$ SINISTRO camera with an image scale of $0\farcs389$ per pixel, resulting in a $26\arcmin\times26\arcmin$ field of view. 

The images were calibrated by the standard LCOGT {\tt BANZAI} pipeline \citep{McCully2018}, and differential photometric data were extracted using {\tt AstroImageJ} \citep{Collins2017}. We used circular photometric apertures with radii $5\farcs1$ to $8\farcs2$. The target star aperture included all of the flux from the nearest known neighbour in the \textit{Gaia} DR3 catalogue (Gaia DR3 1433062331333157632), which is $1\farcs9$ northeast of TOI-1453 and has a delta T-band magnitude of 5.4, making it $\sim$100 times fainter than TOI-1453. 

We visually compared the light curves of the 16 nearby stars within $2\farcm5$ of the target star that are bright enough to potentially cause the TESS detection. No obvious nearby eclipsing binaries (NEBs) were detected in 15 of 16 of the stars, but the $1\farcs9$ neighbour result is inconclusive due to heavy blending with the target star in the photometric aperture. We conclude that the transit is occurring either in TOI-1453, or the $1\farcs9$ neighbour star. We investigate this further in Sect.\,\ref{Sect:Validation}. 
The light curve data are available on the {\tt EXOFOP-TESS} website\footnote{\url{https://exofop.ipac.caltech.edu/tess/target.php?id=198390247}}.

\subsubsection{Maury Lewin Observatory}
A full transit window of TOI-1453.02 was observed from Maury Lewin Observatory in Ic band on UTC 2020 May 09. 
The Maury Lewin Astronomical Observatory 0.36\,m telescope is located near Glendora, CA. It is equipped with a $3326\times2504$ SBIG STF8300M camera having an image scale of $0\farcs84$ per pixel, resulting in a $23\arcmin\times17\arcmin$ field of view. The images were calibrated and differential photometric data were extracted using {\tt AstroImageJ}. The shallow $\sim 250$ ppm transit signal would be difficult to detect from ground-based observatories, so we instead checked for NEBs in the 29 nearby stars within $2\farcm5$ from the target star that are bright enough to potentially cause the TESS detection. We found no obvious NEB signals, but the results do not rule out NEBs at high confidence in most of the stars, including the $1\farcs9$ neighbour \textit{Gaia} DR3 1433062331333157632. The NEB check light curves are available on the {\tt EXOFOP-TESS} website. 

\subsubsection{Adams Observatory}
Another full transit window of TOI-1453.02 was observed from Adams Observatory in the Rc band on UTC 2020 June 04. The Adams Observatory is located at Austin College in Sherman, TX. The 0.6\,m telescope is equipped with a FLI ProLine PL16803 detector that has an image scale of $0\farcs38$ pixel$^{-1}$, resulting in a $26\arcmin\times26\arcmin$ field of view. The observations from this facility were also aimed at ruling out NEBs in stars in close proximity to TOI-1453 at the predicted ephemerides of the candidate planet b. We did not detect any clear NEB signals, although again we cannot rule out this possibility with a high level of confidence. The results are available on the {\tt EXOFOP-TESS} website.

\subsubsection{MuSCAT2}
MuSCAT2 \citep{Narita2019} observed TOI-1453 on the night of March 13, 2024. The instrument is a multi-band optical imager mounted on the 1.5 m Telescopio Carlos S\'{a}nchez (TCS) at the Teide Observatory, Spain. MuSCAT2 is equipped with four cameras capable of taking simultaneous images in the $g'$, $r'$, $i'$, and $z_s$ bands. Each CCD has 1024\,$\times$\,1024 pixels with a field of view of $7.4 \times 7.4$ arcmin$^2$.

The telescope was defocused to avoid saturation of the target. The exposure times were set to 10 s, 10 s, 5 s and 5 s in $g'$, $r'$, $i'$ and $z_s$ respectively. The data were reduced by the MuSCAT2 pipeline \citep{Parviainen2019}; the pipeline performs dark and flat field calibrations, aperture photometry, and transit model fitting including instrumental systematics. Due to the faint reference stars, the light curve scatter was too large to detect the transit with any significance. Our analysis eliminated 3 stars as possible EBs during the observed window.

\subsection{HARPS-N high precision spectroscopy}
All the above follow-up observations increased our confidence in the likely planetary nature of the candidates TOI-1453.01 and TOI-1453.02. Early, we began a precise RV follow-up campaign with HARPS-N \citep{Cosentino2012, Cosentino2014}. HARPS-N is a high-resolution (R = 115\,000) echelle spectrograph mounted on the 3.6m Telescopio Nazionale Galileo (TNG) at Roque de los Muchachos observatory, La Palma, Spain. The HARPS-N instrument is stabilised in temperature and pressure so to achieve the submeter-per-second accuracy and its spectra cover a wavelength range between 383 and 693\,nm. 

\subsubsection{DRS}
We began the observations of TOI-1453 with HARPS-N on 2020 March 12 and gathered spectra for two seasons until 2021 September 19. In total, we obtained 100 RV measurements as part of the HARPS-N Collaboration Time (HCT). Each observation had an exposure time of 1800\,s, which led to a median S/N of 58 in order 50. The RVs were extracted with version 3.0.1 of the data reduction software \citep[DRS,][]{Dumusque2021}, adapted from the ESPRESSO pipeline with additional HARPS-N systematics corrections. We employed a G9 spectral mask to derive the cross-correlation function (CCF) and extract the RVs. Our DRS RV time series presents a median RV uncertainty per data point of 1.56 m/s and a root mean square (rms) deviation of 5.04 m/s. 

\subsubsection{s-BART}
We also extracted the RVs with s-BART \citep{Silva2022}. This is a template-matching pipeline to extract achromatic RV variations. After removing from the spectra wavelength domains sensitive to stellar activity and telluric features, the pipeline creates a spectral template of the star. The latter is built from the series of observations obtained on the given star and assembling the template order-by-order. Finally, it employs a semi-Bayesian approach to undertake the template-matching, assuming each spectrum differs from the template only through a certain Doppler shift. The posterior RV is derived from each observation using a Markov chain Monte Carlo  (MCMC) algorithm. This technique makes it possible to obtain reliable RV uncertainties. The s-BART time series presents a median RV uncertainty of 1.28\,m/s and a rms deviation of 4.03\,m/s. 

\subsubsection{YARARA}
Finally, the RV extraction was performed with YARARA \citep{Cretignier2021}. YARARA is a post-process applied to the spectra derived by the DRS meant to further clean them from instrumental and Earth atmosphere signatures. The spectra are thoroughly corrected for cosmic rays, tellurics, ghosts, fringing, and instrumental focus variations. Additionally, YARARA can also correct for some stellar activity features via a multi-linear regression approach with the S-index, for each wavelength of the spectral time series. 

However, this reduction step did not lead to a smaller scatter in the RV time series of TOI-1453, and no clear activity signal was detected. Therefore, we omitted this correction. 
The merged-order 1d spectra were continuum-normalised with RASSINE \citep{Cretignier2020b}, and the RVs were extracted from the CCF computed on a careful line selection \citep{Cretignier2020a}. Together with the RVs, we also extracted various stellar activity indicators such as the full width at half maximum (FWHM) of the CCF, the \ha, and the \rhk. The YARARA RV time series has a median RV uncertainty of 1.04\,m/s and a rms deviation of 2.73 m/s, which are significantly reduced compared to the DRS due to the powerful correction of additional systematics and improved selection of stellar lines to compute the CCF. 

Because both the RV uncertainties and rms are the smallest with the YARARA time series, we use the latter in the  analyses presented in this paper. Among the 100 observations, 6 were rejected due to anomalous CCFs, leaving us with 94 measurements. The data have been made available via the Centre de Données astronomiques de Strasbourg (CDS) and Table \ref{tab:RV} illustrates the format of this dataset with the five first entries.

% **************************************************** STELLAR PARAMETERS 
\section{Stellar characterisation} \label{Sect:StellarParameters} 
In this section we describe our analyses to characterise TOI-1453. We refer to Table \ref{tab:StellarParam} for the adopted stellar parameters. 

\subsection{Astrometry} \label{Sect:Astrometry}
According to the third \textit{Gaia} data release \citep{GaiaCollab2016, GaiaCollab2023}, the star draws a yearly parallax motion of 12.676$\pm$0.010 milliarcsec. This corresponds to a distance of 78.89$\pm$0.06\,pc from the Sun. Coupling this distance measurement and the systemic RV with the measured proper motions ($\mu_{\alpha}$, $\mu_{\delta}$) = (-70.974, -0.085) mas yr$^{-1}$, we calculated the $U_{LSR}$, $V_{LSR}$, $W_{LSR}$ velocity components of the star in the Galaxy with respect to the Local Standard of Rest (LSR). This estimation includes the correction of the solar motion in the LSR, which we took from \citet{Dehnen1998}. We find reasonably small velocities: ($U_{LSR}$, $V_{LSR}$, $W_{LSR}$) = (-58.99, -5.33, -10.61) km/s. Based on those velocities, we computed the probabilities that TOI-1453 belongs to the thin disc, thick disc and halo following the work of \citet{Reddy2006}. We obtained $P_{thin}$=98.5$\%$, $P_{thick}$=1.5$\%$, $P_{halo}$<0.1$\%$. Our analysis of TOI-1453 kinematics hence indicates that the star belongs to the thin Galactic disc. 

As was shown in Sect. \ref{Sect:Obs:TESS}, TOI-1453 has a close stellar companion with a magnitude difference of $\Delta$\,G = 5.24. They share similar proper motions, according to the \textit{Gaia} data, which suggests that they form a stellar binary system. As a result, TOI-1453.01 and TOI-1453.02 are classified as S-type planet candidates, meaning that they presumably orbit a single component of a multi-star system. The angular separation between the two stellar components is estimated to 1.9$''$, which translates into a projected separation of $\sim$150\,AU given the distance estimated from the \textit{Gaia} parallax. 
In their recent work, \citet{Christian2024} estimated the mass of the stellar companion to TOI-1453, based on an empirical relation between the mass and the magnitude in the K band for low-mass stars \citep{Mann2019}. They found a mass of 0.238$\pm$0.029\,M$_{\odot}$. They also performed a fit of the \textit{Gaia} astrometric data, from which they constrained the semi-major axis of the binary to be 171$\substack{+127 \\ -58}$AU and a (poorly constrained) orbital eccentricity of 0.63$\substack{+0.32 \\ -0.39}$. Additionally, they measured an orbital inclination of 91.98$\substack{+3.86 \\ -1.27}$deg, and concluded that the orbit of the binary is aligned with the planetary orbits. These results have important implications for the disc-binary interactions that took place in this system. In Sect.\,\ref{Sect:Validation}, we discuss our analyses aimed at identifying the true host to the transits. 

Based on the mass and orbital elements of the stellar companion to TOI-1453, we could estimate its expected impact on the RV time series to be $\sim$2.3\,ms$^{-1}$ over the timespan of the HARPS-N observations, under the hypothesis of a circular orbit. We fit the RV data with a linear trend and obtained a slope of -0.37\,ms$^{-1}$/year, i.e. less than 1\,ms$^{-1}$ drift over the timespan of the HARPS-N observations. 
This contrast between the observed and expected drifts is likely due to the large orbital eccentricity which induces asymmetric RV variations. Another possibility is the existence of an underlying additional RV drift that counterbalances the effect of the binary. This additional drift could be instrumental or stellar in origin. As HARPS-N is known to be very stable on such timescales, we rejected the instrumental hypothesis. A drift due to the stellar magnetic cycle would have an opposite direction and similar amplitude to the drift due to the binary. The combination of these conditions makes the stellar origin less likely than that of an eccentric binary orbit. The high eccentricity is supported by the work of \citet{Christian2024} and would create a significantly smaller RV variation along most of the orbit.

\subsection{Stellar atmospheric parameters}

\begin{table}[!h]
\centering
\begin{threeparttable}
\caption{TOI-1453 stellar parameters.}
\label{tab:StellarParam}
\begin{tabular}{@{}llcc@{}}
\toprule
Parameter [units]     &    Value    &   Source\tnote{$\dagger$}    \\ \midrule 
\multicolumn{2}{l}{\textit{Designations and coordinates}} \vspace{0.15cm} \\ 
TOI ID  &  1453  &   \vspace{0.1cm} \\ 
TIC ID  &  198390247  &  \vspace{0.1cm} \\   
2MASS ID & J17125372+5711520 &  \vspace{0.1cm} \\   
\textit{Gaia} DR3 ID & 1433062331332673792 &  \vspace{0.1cm} \\  
RA (J2000) [$h m s$]   & 17:12:53.72  & 1  \vspace{0.1cm}  \\
DEC (J2000) [$d m s$]  &  +57:11:52.00  & 1  \vspace{0.3cm} \\ 
\multicolumn{2}{l}{\textit{Magnitudes and astrometric solution}} \vspace{0.15cm} \\ 
$B$ & 12.03$\pm$0.15 & 2 \vspace{0.1cm} \\ 
$V$ & 11.11$\pm$0.08 & 2 \vspace{0.1cm} \\
$G$ & 10.7279$\pm$0.0028 & 1 \vspace{0.1cm} \\ 
$J$ & 9.275$\pm$0.023 & 3 \vspace{0.1cm} \\ 
$H$ & 8.766$\pm$0.029 & 3 \vspace{0.1cm} \\ 
$K$ & 8.686$\pm$0.019 & 3 \vspace{0.1cm} \\ 
$W1$ & 8.635$\pm$0.023 & 4 \vspace{0.1cm} \\ 
$W2$ & 8.678$\pm$0.02 & 4 \vspace{0.1cm} \\ 
$W3$ & 8.604$\pm$0.021 & 4 \vspace{0.1cm} \\ 
$\mu_{\alpha}$ [mas/yr] & -70.974$\pm$0.013 & 1 \vspace{0.1cm} \\ 
$\mu_{\delta}$ [mas/yr] & -0.085$\pm$0.015 & 1 \vspace{0.1cm} \\ 
Parallax $\pi$ [mas]  & 12.676$\pm$0.010 & 1 \vspace{0.1cm} \\ 
Distance $d$ [pc] & 78.885$\pm$0.064 & 5 \vspace{0.1cm} \\ 
$U_{LSR}$ [km/s]\tnote{1} & -58.99$\pm$0.36  & 5 \vspace{0.1cm} \\ 
$V_{LSR}$ [km/s]\tnote{1} & -5.33$\pm$0.62  & 5 \vspace{0.1cm} \\ 
$W_{LSR}$ [km/s]\tnote{1} & -10.61$\pm$0.59  & 5 \vspace{0.3cm} \\ 
\multicolumn{2}{l}{\textit{Physical parameters}} \vspace{0.15cm} \\ 
\teff [K] &  4975$\pm$68  & 5 \vspace{0.1cm} \\  
$[$Fe/H$]$ [dex] &  -0.31$\pm$0.06  &  5  \vspace{0.1cm} \\  
$\log g_{iso}$ [cgs] & 4.576$\pm$0.015 &  5  \vspace{0.1cm} \\ 
$v \sin i$ [km/s]  &  $<$ 2  &  5  \vspace{0.1cm} \\  
\rhk & -4.91$\pm$0.06 & 5  \vspace{0.1cm} \\  
Mass [$M_{\odot}$]\tnote{2}  &  0.715$\pm$0.035  &  5  \vspace{0.1cm} \\  
Radius [$R_{\odot}$]\tnote{2}  &  0.720$\pm$0.029  &  5  \vspace{0.1cm} \\  
$\rho_{\star}$ [$\rho_{\odot}$]  &  1.918$\substack{+0.076 \\ -0.034}$  &  5  \vspace{0.1cm} \\ 
Age [Gyr]  &  12.0$\substack{+1.1 \\ -3.8}$  &  5  \vspace{0.1cm} \\  \bottomrule 
\bottomrule
\end{tabular}
\begin{tablenotes}
\item[1] The uncertainties reported on $U_{LSR}$, $V_{LSR}$, $W_{LSR}$ account for the uncertainties on the motion of the Sun in the LSR reported by \citet{Dehnen1998}.
\item[2] The uncertainties on the stellar mass and radius account for the systematic differences between various stellar evolution models. 
\item[$\dagger$] \textbf{References.} 1. \textit{Gaia} DR3 \citep{GaiaCollab2023}   -  2. Tycho-2 \citep{Hog2000}  -  3. 2MASS \citep{Cutri2003}  -  4. WISE \citep{Cutri2012}  -  5. This work 
\end{tablenotes}
\end{threeparttable}
\end{table}

The high-resolution HARPS-N spectra were used to constrain various stellar parameters. To proceed, we employed three independent characterisation techniques: SPC, ARES+MOOG, and CCFPams. Ultimately, we combined the results from the three methods in order to obtain final stellar parameters that account for model systematics. This process of stellar characterization was described in detail in \citet{Mortier2020}. Here we briefly synthesise the three methods. 

The Stellar Parameter Classification tool \citep[SPC,][]{Buchhave2012, Buchhave2014} is a spectrum synthesis method that works on the individual spectra and merges the results into a weighted average. Each HARPS-N spectrum is cross-correlated with a library of model template spectra. The stellar parameters for TOI-1453 derived by SPC are \teff\,=\,5044$\pm$50\,K, \logg\,=\,4.68$\pm$0.1, [m/H]\,=-0.21$\pm$0.08. Additionally, the SPC method constrains the projected rotational velocity of the star to v\,$\sin$i\,<\,2\,km/s. 

The ARES+MOOG method is a curve-of-growth technique based on a line list of neutral and ionised iron lines \citep{Sousa2014}. The equivalent widths of those lines were computed on the HARPS-N stacked spectra using \texttt{ARESv2} \citep{Sousa2015}\footnote{\url{https://github.com/sousasag/ARES}}. These results are then injected into the radiative transfer code \texttt{MOOG} to extract parameters such as \teff, $\log~g$ and iron abundance [Fe/H]. We note that the HARPS-N wavelength range contains a large number of iron lines. Therefore, we consider the measurement of [Fe/H] to be equivalent to the overall metallicity of the star, [m/H]. Following this procedure, we obtained \teff\,=\,4932$\pm$86\,K, \logg\,=\,4.87$\pm$0.16, [Fe/H]\,=-0.35$\pm$0.05. 

Finally, CCFPams \citep{Malavolta2017}\footnote{\url{https://github.com/LucaMalavolta/CCFpams}} estimates \teff, $\log~g$ and the iron abundance from the measurement of the equivalent width of the CCFs, using an empirical calibration relation. This technique provides \teff\,=\,4948$\pm$68\,K, \logg\,=\,4.693$\pm$0.228, [Fe/H]\,=-0.35$\pm$0.05. The surface gravity that we report includes a correction factor coming from a survey of transit-constrained $\log~g$, as presented by \citet{Mortier2014}. 

Noteworthy, we found interstellar absorption in the HARPS-N spectra in the wide NaD lines around 5890 and 5896 ${\AA}$. The absorption features are consistently blue-shifted by $\sim$ 15.5 km\,s$^{-1}$ with respect to the centre of the lines. We provide further details on this observation in Fig. \ref{Fig:NaI_absorption}.

\subsection{Stellar radius, mass, and age}
We ran the \texttt{isochrones} code \citep{Morton2015} with the MESA Isochrones and Stellar Tracks set of isochrones \citep[MIST,][]{Dotter2016} to estimate the stellar mass, radius, and age. The input parameters for MIST were the effective temperature and metallicity, the \textit{Gaia} parallax, as well as the stellar apparent magnitude in eight photometric bands: $B$, $V$, $J$, $H$, $K$, $W1$, $W2$, and $W3$. We ran the MIST code three times, for each of the three input \teff \, and [Fe/H], with the purpose of alleviating systematic differences in stellar modelling \citep[e.g.][]{Tayar2022}. Finally, we combined the three posteriors and report the results in Table \ref{tab:StellarParam}. 
For the stellar mass and radius, we add a relative uncertainty of 4$\%$ in quadrature to the uncertainties reported by our isochrones fits, in order to account for the systematic differences between various stellar evolution models \citep{Tayar2022}. 

 In this table, we report the value for \logg \, from our isochrones fits, however, we note that \teff \, and [Fe/H] are better constrained from our spectroscopic analyses. For these two parameters, we report the weighted averages from the three methods: \teff =4975$\pm$68\,K, [Fe/H]=-0.31$\pm$0.06. The results indicate that TOI-1453 is slightly metal-poor, while from our isochrones fits we estimate the age to be 12$\substack{+1.1 \\ -3.8}$ Gyrs. Those results have to be considered in regards of the kinematics of the star in the Galaxy, which unambiguously revealed that the star belongs to the thin disc (cf. Sect. \ref{Sect:Astrometry}). 
In a recent work, \citet{Gallart2024} reconstructed the star formation history in the solar neighbourhood within 100\,pc, using \textit{Gaia} data. They showed that at this location, the disc of the Milky Way started to form $\sim$11\,Gyr ago at roughly solar metallicity, and that the metallicity of the stellar population followed a slightly decreasing trend until $\sim$6\,Gyr ([m/H] in-between [-0.5,0]). The metallicity, age, and kinematics estimations of TOI-1453 are in agreement with this global picture.

\subsection{Stellar activity} \label{Sect:StellarActivity}
The HARPS-N spectra provide us with valuable information about the level of activity of TOI-1453 during the observations. We extracted the YARARA \rhk time series. To proceed, we fitted a profile of plage on the CaII\,H and K lines \citep[see][]{Cretignier2024} but adapted for the K spectral type \citep[see Fig.10 in][]{Cretignier2024b}, which leads to a precise but inaccurate S-index time series. Then, we calibrated this time series via a linear transformation on the \rhk estimated from the DRS and our log S-index. Finally, we used the coefficients of this transformation to build a precise and accurate YARARA \rhk time series. The latter indicates \rhk =-4.91$\pm$0.06, pointing towards a low activity level. We applied the empirical relation of \citet{Noyes1984} to obtain a rough estimate on the stellar rotation period. We obtained $P_{rot}$=42.9 days,  while we undertook the same process for the relation of \citet{Mamajek2008} and found $P_{rot}$=43.8 days. 

\begin{figure*}
    \centering
\includegraphics[width=\textwidth]{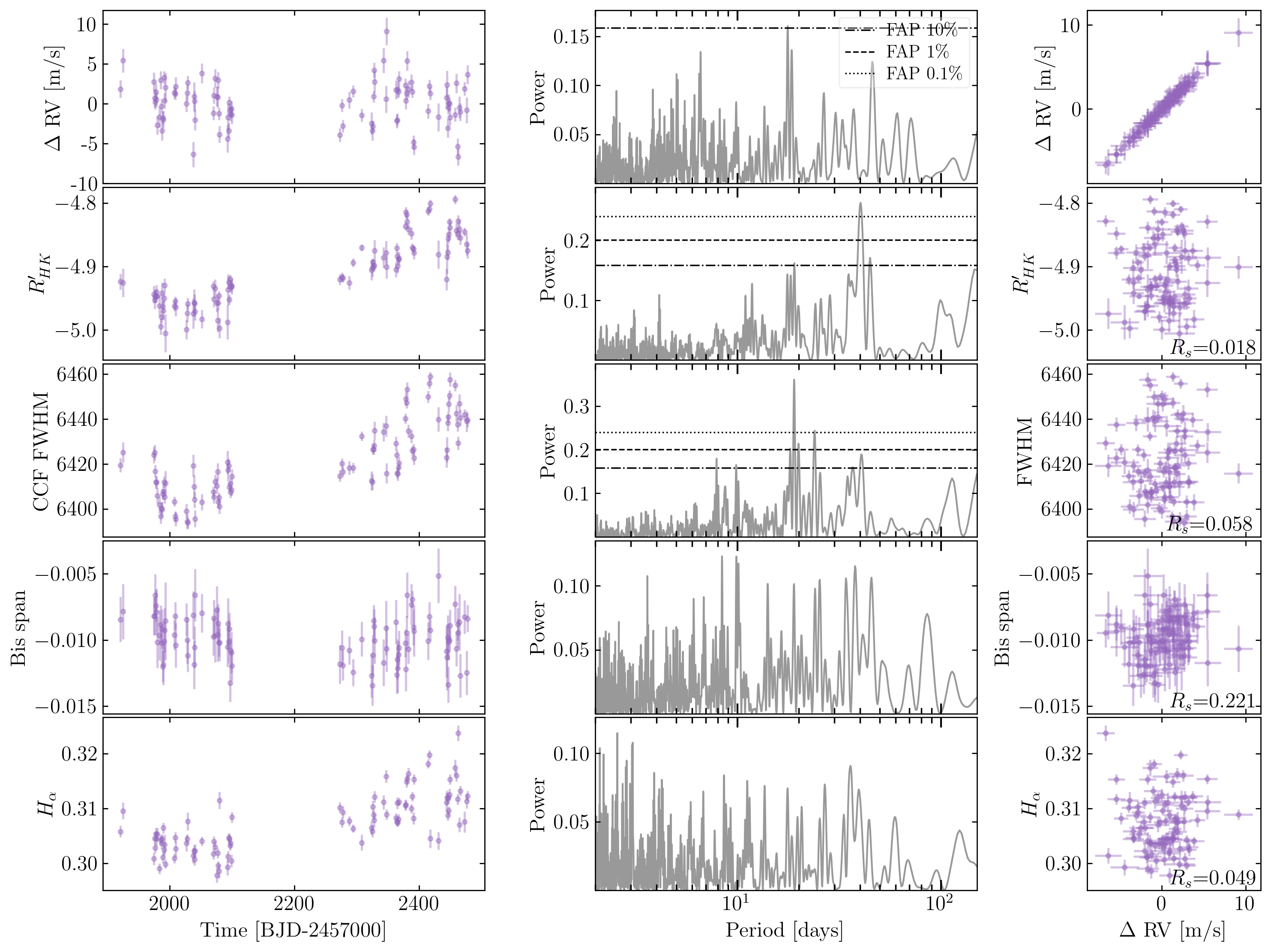}
\caption{HARPS-N YARARA RVs, \rhk, CCF FWHM, bisector span, and \ha time series. \textit{Left}: The various time series. \textit{Middle}:  GLS periodograms, together with the 10, 1, and 0.1$\%$ FAP levels. \textit{Right}:  Correlation plots with the RVs. In the plots, we also indicate the Spearman correlation coefficient.} 
\label{Fig:Spectro_AllInfo} 
\end{figure*}

We report in Fig.\,\ref{Fig:Spectro_AllInfo} the time series, periodograms and correlation plots of the YARARA RVs together with a few spectroscopic activity indicators: \rhk, the FWHM and bisector span of the CCF, and \ha. The plots reveal little to no correlation between our RV time series and the activity indicators. Additionally, we observe that the RV time series is contained within $\pm$ 10 ms$^{-1}$. To compute the periodogram of \rhk, FWHM and \ha, we first subtracted a quadratic trend from the time series. The concave shape of these trends suggests a phase of low stellar magnetic activity. We observe a significant signal at 40.3 days in the residual periodogram of \rhk, and 19.0 days in the periodogram of FWHM. The strongest (but not significant) periodicity in the RVs is 17.7 days. 

In an attempt to further constrain stellar rotation in the HARPS-N data, we used \texttt{SPLEAF} \citep{Delisle2020, Delisle2022} to fit a multi-dimensional Gaussian Process (GP) simultaneously on the RVs and \rhk. To proceed, we employed the Exponential-sine periodic (ESP) kernel, which is an approximation of the squared-exponential periodic kernel. It is an efficient kernel to capture stellar rotation features \citep[e.g.][]{Stalport2023}. We fitted the GP together with a Keplerian around P=6.59 days to account for the signal of the outer planet candidate (there is no serious hint of the inner candidate in the RV data alone). However, we could not reach a satisfying convergence despite a vast exploration of different priors. While the period of the GP consistently peaked at $\sim$39 days, the other hyperparameters were strongly unconstrained. We also tried to focus the analyses only on the second season of observations, which shows a slightly higher activity level. The results were similarly inconclusive. We also tested the multi-dimensional GP on FWHM, bisector span, and \ha, all leading to bad convergence. Finally, we tested the simpler case of a GP applied on the RV only, which might ease the convergence when stellar activity is weak. Again, we fitted this GP simultaneously with one Keplerian. We were unsuccessful in constraining the hyperparameters, this time including also the period of the GP. These results confirm the low activity of the star during our HARPS-N observations.

\begin{figure}
    \centering
\includegraphics[width=\columnwidth]{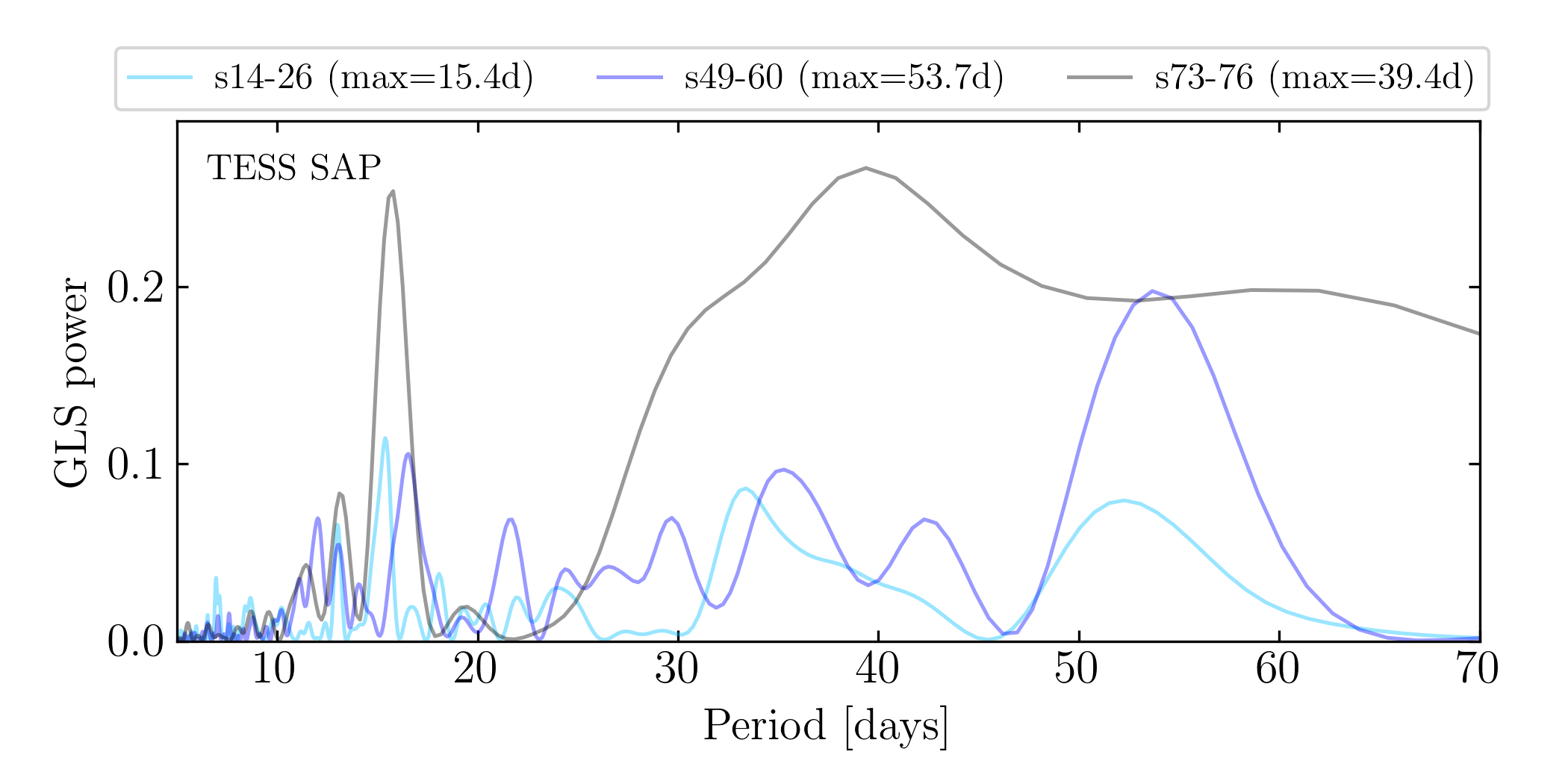}
\caption{GLS periodograms of the TESS SAP photometry separated into three datasets: sectors 14 to 26 (light blue), sectors 49 to 60 (dark blue), and sectors 73 to 76 (gray). In the legend, the periods at which the power is maximum are indicated inside parentheses for each dataset.} 
\label{Fig:TessSAP_Periodogram} 
\end{figure}

We analysed the TESS light curve in search for periodicity in the photometry. To proceed, we downloaded the SAP fluxes and removed the flux measurements strongly contaminated by the Earth or Moon scattered light. Thanks to the extensive timespan of the TESS observations, we defined three datasets roughly corresponding to the three cycles of observations: sectors 14 to 26 (12 sectors), sectors 49 to 60 (9 sectors), and sectors 73 to 76 (4 sectors). For each dataset, we undertook a period search by computing generalised Lomb-Scargle (GLS) periodograms. We present the results in Fig. \ref{Fig:TessSAP_Periodogram}. This analysis shows an increasing activity over the timespan of the observations, as is also hinted in the \rhk\  time series from the HARPS-N observations. Beyond this trend, we have not observed any significant periodicity. Noteworthy, the main periodicity in the last cycle of TESS data peaks at 39.4d, similar to the period found in the \rhk. However, more than 800 days separate these two datasets, so this should not help our multi-dimensional GP converge. 

Our analyses suggest a tentative stellar rotation period around 40 days. However, we remind that this result is based on observations taken during a phase of low stellar activity, which affects the significance of our conclusion.

% **************************************************** DATA ANALYSIS 
\section{Data analysis} \label{Sect:DataAnalysis} 
\subsection{Photometric analyses} \label{Sect:TransitFit}
We decided to utilise exclusively the TESS data for our photometric analyses, and to not include the ground-based LCOGT photometry or the other photometric datasets listed in Sect. \ref{Sect:GroundBasedPhot}. The reason is that the planetary transits in the ground-based data are very shallow (the inner planet is not detected, the outer planet is at the limit of detection). Therefore, these data do not help constrain the transits shape and depth. Adding such ground-based photometry can still be very useful when it significantly extends the observational baseline of the system, which can serve to refine the orbital periods. However in our case, the ground-based photometry does not extend the timespan of the extensive TESS dataset. 

We started our planet analyses by searching for the transit signals in our extracted TESS light curves. Using the Python package Transit Least Squares \citep[TLS,][]{Hippke2019}, we consistently found the transit signals of the two planet candidates throughout the TESS observations. The linear ephemerides obtained with this preliminary search were then used as priors to fit a two-planet model to the light curve with \texttt{Juliet}. \texttt{Juliet} models photometric data with the \texttt{batman} package \citep{batman}. Our model consists of the sum of two transit components and a noise component. In our case, the noise component simply consists of a jitter term $\sigma_{TESS}$ added in quadrature to the photometric measurements (this noise model is satisfactory as we explained in Sect. \ref{Sect:Obs:TESS}). 

The transit model for each planet is parametrised with the orbital period, $P$, the reference time at inferior conjunction, $T_0$, the orbital eccentricity, $e$, the argument of periastron, $\omega$, the relative radius, $\left(R_p/R_{\star}\right),$ and the impact parameter, $b$. The latter two parameters were included via the ($r1$, $r2$) parametrisation suggested by \citet{Espinoza2018}, and with uniform priors in [0,1]. We fixed $e$ to $0$, and set Gaussian priors on $P$ and $T_0$ based on the results of TLS (for planet b: $\mathcal{N}$(4.31, 0.1) and $\mathcal{N}$(2459392.91, 0.1), for planet c: $\mathcal{N}$(6.59, 0.1) and $\mathcal{N}$(2459395.37, 0.1), where $\mathcal{N}$ indicates a normal distribution). In this transit model, the stellar density $\rho$ is used together with $P$ to constrain the scaled semi-major axis $a/R_{\star}$ of each planet, while reducing the number of parameters - from the two parameters $a_b/R_{\star}$ and $a_c/R_{\star}$, we now only fit for the stellar density \citep{juliet}. We set priors on $\rho$ according to our stellar characterisation analyses -- cf. Sect. \ref{Sect:StellarParameters}. The transit model also contains instrument-dependent parameters such as an out-of-transit flux offset, $\delta_{flux}$ (prior: $\mathcal{N}$(0, 0.1)), a dilution parameter, $D$, and the limb darkening (LD) coefficients. The dilution parameter accounts for the light contamination from the neighbouring stars, which results in an underestimation of the transit depth. This parameter is defined in \texttt{Juliet} as: 
$$ D = \dfrac{1}{1 + \sum_{k} F_k/F_T}, $$
where $F_k$ is the flux of each star $k$ in the photometric aperture except the target star, whose flux is $F_T$. 
Given that we used the same aperture than the SAP and PDCSAP to extract the light curve, the dilution factor is simply given by the \texttt{crowdsap} parameter indicated in the FITS files of the TPF. This metric provides the ratio between the target flux and the total flux inside the aperture, which is equivalent to $D$. Its average value over all the sectors is $0.98435$. To account for the effect of LD, we used a quadratic LD law with coefficients $q_1$ and $q_2$ defined in \citet{Kipping2013}. The stellar parameters were used to constrain those coefficients for the TESS passband with the help of \texttt{LDCU}\footnote{\url{https://github.com/delinea/LDCU}}. \texttt{LDCU} is a modified version of the Python routine implemented by \citet{Espinoza2015} that computes the LD coefficients and their corresponding uncertainties using a set of stellar intensity profiles accounting for the uncertainties on the stellar parameters \citep[e.g.][]{Deline2022}. The stellar intensity profiles are generated based on two libraries of synthetic stellar spectra: ATLAS \citep{Kurucz1979} and PHOENIX \citep{Husser2013}. The code takes as input various stellar parameters $T_{\rm eff}$, $\log{g}$,$[m/H]$, which we estimated in Sect. \ref{Sect:StellarParameters}. For each intensity profile, \texttt{LDCU} performs three fits for different atmosphere model data selection processes (all, limb-excluded, and interpolated data). To obtain the most conservative LD estimations, we extracted the merged distributions from all the intensity profiles, and further increased the uncertainties by a factor $\sim$3. We used these results to set Gaussian priors on the LD coefficients ($\mathcal{N}$(0.3911, 0.05) and $\mathcal{N}$(0.3358, 0.05) for $q_1$ and $q_2$, respectively). 

\begin{figure*} 
    \centering
    \subfloat[]{\includegraphics[width=\textwidth]{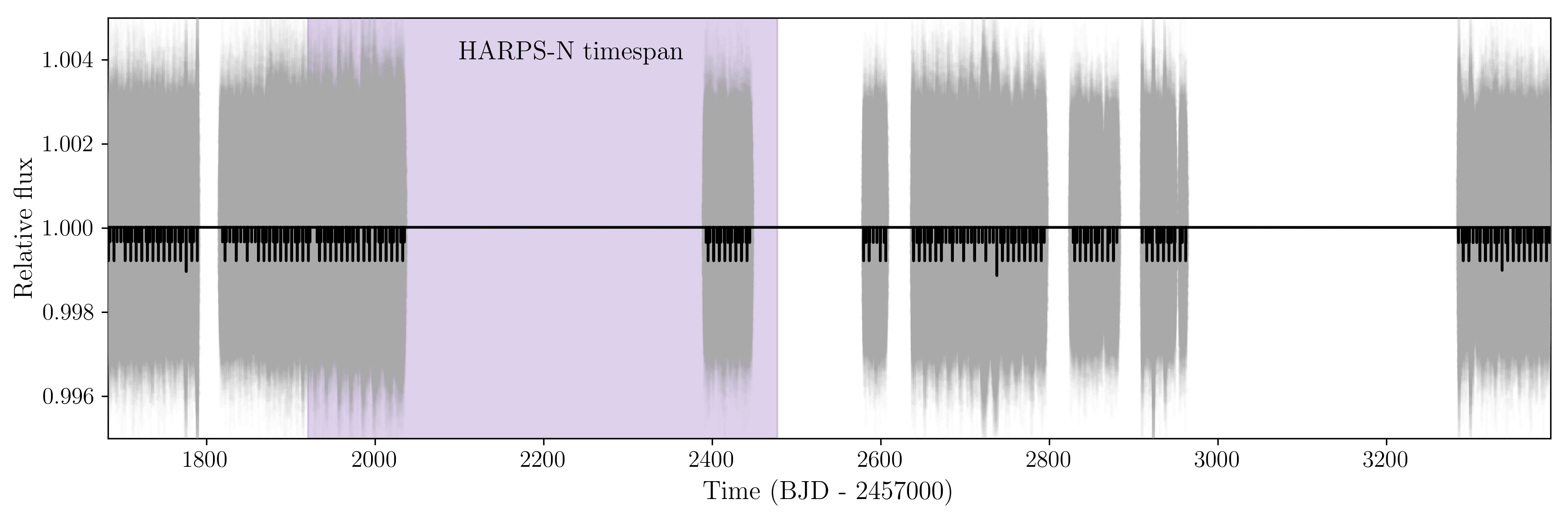}} 
    \par\medskip
    \begin{minipage}{.5\linewidth}
    \centering
    \subfloat[]{\includegraphics[width=\columnwidth]{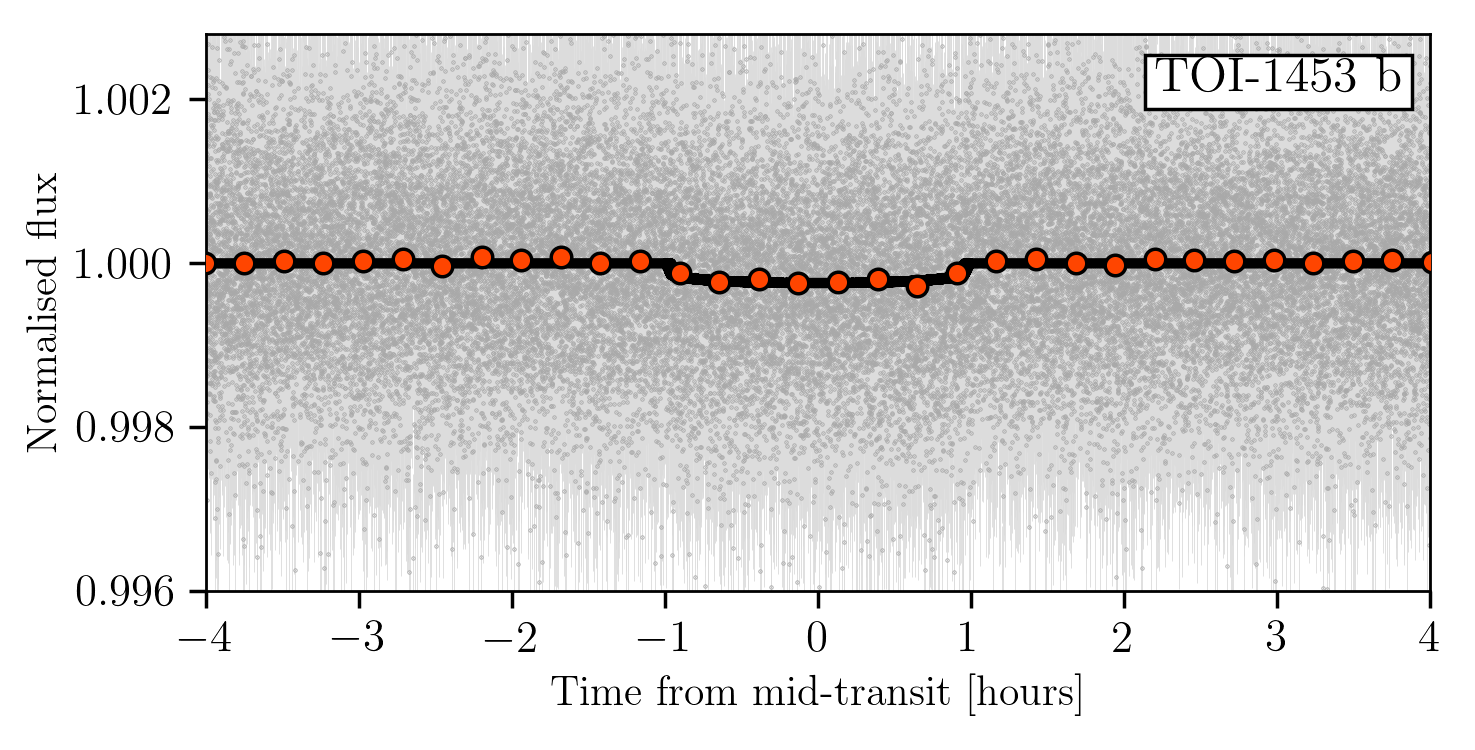}}
    \end{minipage}%
    \begin{minipage}{.5\linewidth}
    \centering
    \subfloat[]{\includegraphics[width=\columnwidth]{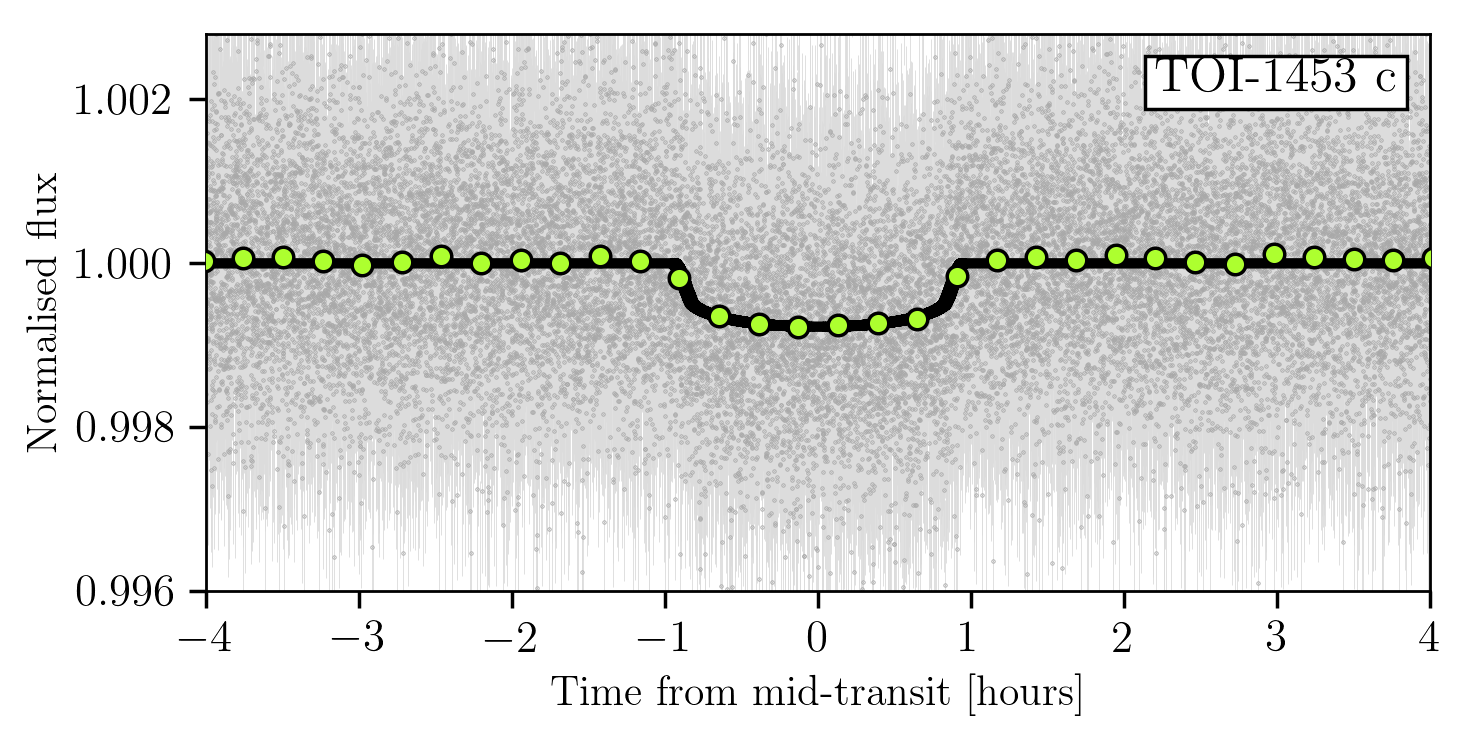}}
    \end{minipage}%
\caption{TESS light curve of TOI-1453 (29 sectors comprised between sector 14 and sector 76) together with the transit model in black. 
(a) Full TESS light curve. The zone highlighted in purple illustrates the timespan of the HARPS-N observations. The three deeper transit features in the model correspond to simultaneous transits of planets b and c. (b) TESS light curve phase-folded on the transits of TOI-1453\,b (TOI-1453.02). The orange dots are binned TESS data for better visibility. (c) Same for TOI-1453\,c (TOI-1453.01), where the green dots are binned TESS data.} 
\label{Fig:LCresult} 
\end{figure*}

Having defined the model, we proceeded to fit the data from 29 TESS sectors. The model parameters were explored with the nested sampling algorithm \texttt{dynesty} using a number of live points $n_{\rm live} \geq N^2$, where $N$ is the number of model parameters to be explored (not accounting for the fixed parameters). We assumed that the fit had converged when the variation in log-evidence from one iteration to the next went below 0.01. In Fig. \ref{Fig:LCresult}, we present the TESS light curve together with the best-fit transit model. 

\subsection{Validation of the planet candidates} \label{Sect:Validation} 
To validate the two planets as transiting TOI-1453, we explored all the possible scenarios that could create our observations. To proceed, we used \texttt{TRICERATOPS}\footnote{\url{https://github.com/stevengiacalone/triceratops}} \citep{Giacalone2020, Giacalone2021}. This publicly available code employs a Bayesian probabilistic approach to test the astrophysical false positive hypotheses such as nearby eclipsing binaries. \texttt{TRICERATOPS} takes as input the normalised light curve phase-folded onto the transit feature of interest. We phase-folded the entire TESS light curve onto the period and phase of TOI-1453.01, and binned the resulting data into 400 data points to increase the computation speed. Additionally, we also provided the contrast curve of the speckle image in the 562\,nm narrow photometric band from Gemini, which helps constrain the false positive probabilities (FPPs). We ran the \texttt{TRICERATOPS} test for TOI-1453.01, and obtained FPP=0.0023$\pm$0.0017. We repeated the same process for TOI-1453.02, and found FPP=0.0211$\pm$0.0020. While this test validates TOI-1453.01 as a planet orbiting TOI-1453, namely TOI-1453\,c, the result is not so clear for TOI-1453.02 since the FPP is slightly above 1$\%$. In particular, we cannot firmly rule out the nearby stellar companion 1.9$''$ away as the source of the transit. We note however that this scenario seems unlikely, because that would imply a transit depth of $\sim$10$\%$ on this faint star and subsequently a large planet around a small star. In contrast, the scenario of two small planets around TOI-1453 is more likely. As was already noted in the context of the Kepler space mission, the probability of FP decreases sharply in multi-planet candidate systems compared to singles \citep{Lissauer2012, Lissauer2014}. These studies revealed a correction factor larger than 10 on the FPPs between single and multi-planet candidate systems. Taking the system multiplicity into account in our FPP estimates above, we conclude that both planet candidates present FPP below 1$\%$. 

To proceed further, we performed a second validation test based on the computation of the stellar density. In Table\,\ref{tab:StellarParam}, we report our estimation of $\rho_{\star}$ based on our fits of isochrones constrained from the HARPS-N spectra (cf. Sect.\,\ref{Sect:StellarParameters}). Additionally, the transit shape independently constrains the stellar density as well. The comparison between the transit-based and the model-based stellar densities represents a good test of the stellar host. The stellar companion to TOI-1453, DR3 1433062331333157632, is much fainter than TOI-1453. Provided they both are at the same distance from the Sun, this implies that the spectral types of the two stars are also very distinct. From the mass estimate of \citet{Christian2024}, $M_B=0.24M_{\odot}$, we expect an effective temperature \teff<3500$K$. At such small \teff, the stellar density is expected to be larger than TOI-1453 by a factor of at least 5 \citep{Baraffe2015}. The densities of the two stars being significantly different from each other, their comparison to the transit-based densities could bring valuable information. 

The stellar density can be derived from the transit parameters following the formula \citep{Seager2003}: 
$$ \rho_{\star} ~ = ~ \left(\dfrac{4~\pi^2}{P^2~G}\right) \left[\dfrac{\left(1+\dfrac{R_p}{R_{\star}}\right)^2 ~ - ~ b^2 ~ (1-\sin^2(T_{14}\pi/P))}{\sin^2(T_{14}\pi/P)}\right]^{3/2}, $$
where $P$ is the orbital period of the considered planet, $R_p$ and $R_{\star}$ are the planetary and stellar radii, $b$ is the transit impact parameter, and $T_{14}$ is the total transit duration. Therefore, we performed a fit unconstrained by the stellar priors so to measure the stellar density from the transit parameters. We set wide uninformative priors on the stellar density and LD coefficients ($\log\mathcal{U}$[500, 10000] kg\,m$^{-3}$ and $\mathcal{U}$[0,1] for $\rho_{\star}$ and LD coefficients $q_1$, $q_2$, respectively), and fixed the dilution factor $D$ to 1. From the posteriors of that analysis, we computed the stellar densities based on the transit parameters of TOI-1453.01 and TOI-1453.02, respectively. We found $\rho_{\star}$(.01)=1.04$\substack{+1.13 \\ -0.84}$ $\rho_{\odot}$ and $\rho_{\star}$(.02)=1.94$\substack{+1.66 \\ -1.61}$ $\rho_{\odot}$. These two stellar densities are compatible with the model-based stellar density of TOI-1453 within 1$\sigma$. Furthermore, they are not compatible with the expected high density of the stellar companion DR3 1433062331333157632, which validates both planets. In the following, we will refer to TOI-1453.02 as TOI-1453\,b, and TOI-1453.01 as TOI-1453\,c.

\subsection{Joint TESS - HARPS-N analysis} \label{Sect:JointFit}

\begin{figure*} 
    \centering
    \includegraphics[width=\textwidth]{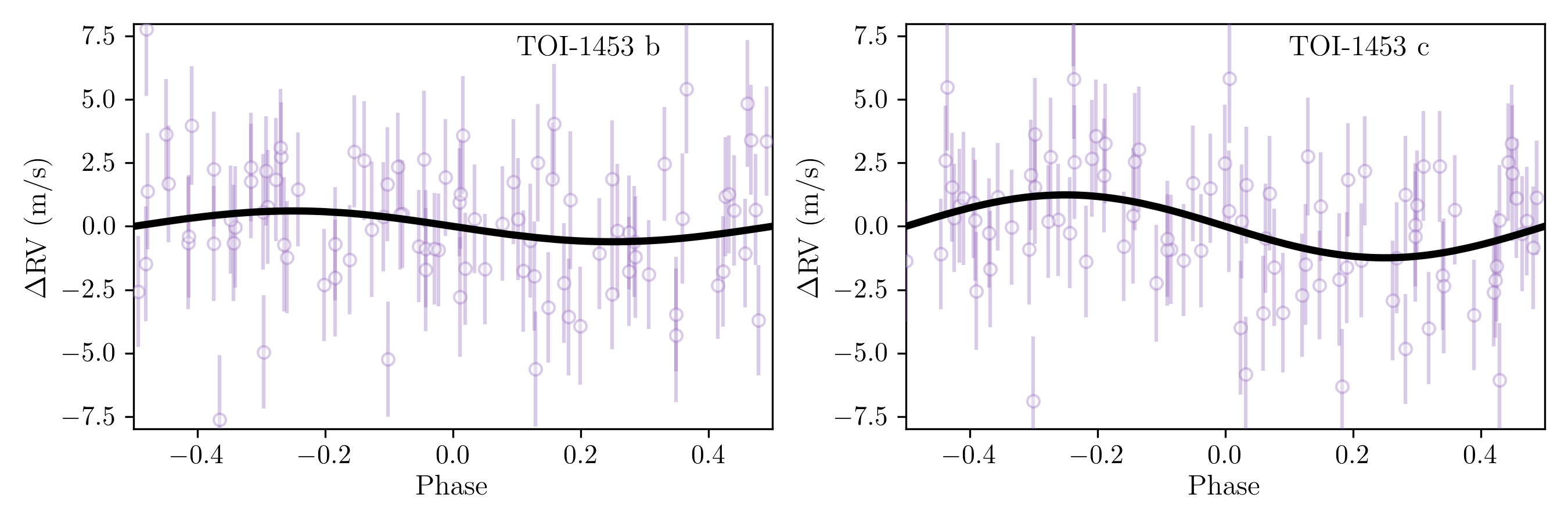}
\caption{HARPS-N RV phase-folded on the results of the joint fit for both TOI-1453 b (left) and TOI-1453 c (right). The black line represents the best-fit model RV curve of each planet.} 
\label{Fig:RVresult} 
\end{figure*}

The results of the transit fit described in Sect. \ref{Sect:TransitFit} were used to constrain a global fit to the transits of each planet and the RVs. Notably, we narrowed down the priors on $P$ and $T_0$, and restricted the relative radius exploration of both planets to values within the range $R_p/R_{\star}$=[0.01, 0.05] so as to reduce the computation cost\footnote{This restricted exploration under the ($r_1$,$r_2$) parametrisation was ensured by setting the following arguments to the \texttt{Juliet} function \texttt{dataset.fit}: rl=0.01, ru=0.05.}. Regarding the RVs, we included the YARARA data presented in Fig. \ref{Fig:Spectro_AllInfo}. The addition of the RV data leads to the inclusion of four more parameters in the global model: the RV semi-amplitude $K_b$ and $K_c$ of planets b and c, an RV offset $\mu_{HARPSN}$ and a jitter term $\sigma_{HARPSN}$ for the HARPS-N data. We placed wide priors on these parameters. We also tested the inclusion of a RV linear drift in the model, but both slope and intercept were compatible with zero so we decided to not include this trend in the final fit. The set of parameters and priors used in our joint fit is described in Table \ref{tab:PlanetParam}. 

While GP detrending of YARARA RVs can efficiently account for stellar red noise \citep[e.g.][]{Stalport2023}, our observations do not reveal significant stellar activity, as already discussed in Sect. \ref{Sect:StellarActivity}. To further test whether GP detrending was needed, we compared joint transit+RV fits with different input RV data using \texttt{Juliet}. 
In one case, we input the original YARARA RVs, in the other case we use the RVs detrended from the multi-dimensional GP (RV, \rhk). We employed the same two-planet model in both fits with uniform priors on the RV semi-amplitudes between 0 and 100 m/s, and we compared the posterior planet parameters after a Bayesian exploration of the parameter space with \texttt{dynesty} (to decrease the computation time, we performed those analyses on a reduced TESS dataset composed of the sectors 24, 25, 26, 40, 41). The RV semi-amplitudes have negligible differences between the two fits, with $\Delta K \sim$ 0.02 m/s for both planets b and c. Therefore, there is no measurable impact of our GP correction on the planet masses. 
Additionally, we obtain slightly smaller uncertainties on $K_b$ and $K_c$, and hence higher-significance detections, for the fits that use the original YARARA datasets without GP detrending. 
In conclusion, for the global joint transit+RV fit we opt for the original YARARA RVs, without further detrending. 

We undertook a joint fit of the HARPS-N RVs and 29 TESS sectors. We explored the model parameters with \texttt{dynesty} using the same constraint on the number of live points ($n_{\rm live} \geq N^2$) and the same criterion on convergence than with the fit on the photometric data alone. The results of our global joint fit are reported in Table \ref{tab:PlanetParam}. We obtain nearly identical orbital inclinations between both planets. This is an expected outcome of tidal dissipation, which aligns the planetary orbits with the stellar spin given the stellar rotation period is not small compared to the orbital periods \citep[e.g.][]{Bolmont2014, Ogilvie2020}. Therefore, the observed orbital alignment further supports both planets orbiting the same star. Regarding LD coefficients, the posteriors are essentially identical to the priors, despite the conservative approach from \texttt{LDCU}. This highlights here that the LD coefficients are essentially constrained from the stellar parameters rather than the transits.

Thanks to the extensive TESS dataset, we constrain well the transit depths of the inner and outer planets to 224$\pm$12.5\,ppm (2.8$\%$ uncertainty on $R_b/R_{\star}$) and 802$\pm$20.7\,ppm (1.3$\%$ uncertainty on $R_c/R_{\star}$), respectively. With our set of HARPS-N observations, we measure RV semi-amplitudes of $K_b$=0.61\,m/s (with 1.9$\sigma$ significance), $K_c$=1.26\,m/s (non-zero with 3.5$\sigma$ significance), respectively. We present the best fit RV curves of TOI-1453\,b and c in Fig.\,\ref{Fig:RVresult}. Therefore, the RV detection of the inner planet remains insignificant with our current dataset. However, we set reliable constraints on the mass of the outer planet: $M_c$=2.95$\substack{+0.83 \\ -0.84}$\,$M_{\oplus}$. With an estimated radius of 2.224$\pm$0.095\,$R_{\oplus}$, we conclude that this sub-Neptune presents a very low bulk density of 0.267$\substack{+0.087 \\ -0.080}$\,$\rho_{\oplus}$.

\section{Dynamical state and TTVs} \label{Sect:DynamicalState}
\subsection{Proximity of the planets to the 3:2 MMR} \label{Sect:MMR}
The planets, with their orbital periods of $P_b$=4.314 days and $P_c$=6.589 days, present a period ratio of $P_c$/$P_b$=1.527, close to the exact 3:2 commensurability. Therefore, we carried out a dynamical analysis to assert if the planets are dynamically bound into a state of  mean motion resonance (MMR). Furthermore, with these analyses we also investigated the stability of the planetary system. 
We did not include the stellar companion to TOI-1453 in our dynamical analyses. Indeed, our investigations focus on short-term dynamics (and short-term instability), enough to reveal MMR. Within these short timescales, the stellar companion only has a marginal effect on the MMR structure \citep[e.g.][]{Stalport2022b}. 

Chaos maps are powerful tools to visualise MMR in a planetary system \citep[e.g.][]{Gozdziewski2001, Correia2009, Stalport2022b, Daquin2023}. These are 2-dimensional sections of the parameter space, inside which the dynamical chaos is evaluated. We focused our investigations on the short-term chaos which is a direct outcome of the interactions between the MMR \citep{Chirikov1979}. To estimate the amount of chaos in a planetary system, we coupled the results of numerical evolution with the Numerical Analysis of Fundamental Frequencies (NAFF) chaos indicator \citep[][]{Laskar1990, Laskar1993}. NAFF is a process used to precisely compute the orbital mean motion of a planet based on a time series of its mean-longitudes over time. From this time series, a refined Fourier frequency analysis is performed in order to identify the main frequency of the orbital evolution, which is associated with the mean motion. Separating the mean-longitudes time series in two halves, we can apply this technique on each half of the integration and compute the quantity: 
$$\mathrm{NAFF} ~ = ~ \max_{j} ~ \left[ \log_{10} \dfrac{\mid n_{j,2} - n_{j,1} \mid}{n_{j,0}} \right], $$ 
where $n_{j,1}$ and $n_{j,2}$ are the mean motions of planet $j$ over the first and second half of the integration, respectively, and $n_0$ is the initial mean motion of that same planet and serves as a normalisation factor. 
The temporal variation of the mean motion offers direct information on the amount of chaos in the planetary orbit. Indeed, the classical (non-chaotic) theory of dynamics of Laplace and Lagrange predicts constant orbital mean motion over secular timescales. This is no longer true in the presence of chaos. Therefore, the drift in the orbital mean motion $n_{j,2}-n_{j,1}$ is a measure of chaos in a planetary orbit. We estimated the chaos in each of the planetary orbits, $j$, and defined the level of chaos in the whole system as the level of chaos of the planetary orbit that displays the highest degree of variation. 

In practice, we generated a grid of 201$\times$201 systems exploring the initial period ratio $P_c/P_b$ and the orbital eccentricity $e_c$ of the outer planet, while all the other parameters had their initial conditions fixed to the best-fit solution (cf. Table \ref{tab:PlanetParam}). We simulated the evolution of each system in the grid with the adaptive time-step high-order integrator \texttt{IAS15} \citep{Rein2015} included in the Python package \texttt{REBOUND}\footnote{\url{https://rebound.readthedocs.io/en/latest/}} \citep{Rein2012}. The timespan of each simulation was set to 10\,kyr (i.e. about 554000 revolutions of planet c). We included an orbital correction due to General Relativity (GR) effects following the framework of \citet{Anderson1975} and implemented in \texttt{REBOUNDx}\footnote{\url{https://reboundx.readthedocs.io/en/latest/}} \citep{Tamayo2020}. During these simulations, the planet mean longitudes were evenly recorded to build a time series of twenty thousand values. These time series were split in two halves, from which we computed the mean motion of TOI-1453\,b and c on each half and derived the NAFF indicator. This process was repeated for each of the 40401 system configurations. 

\begin{figure} 
    \centering
    \includegraphics[width=\columnwidth]{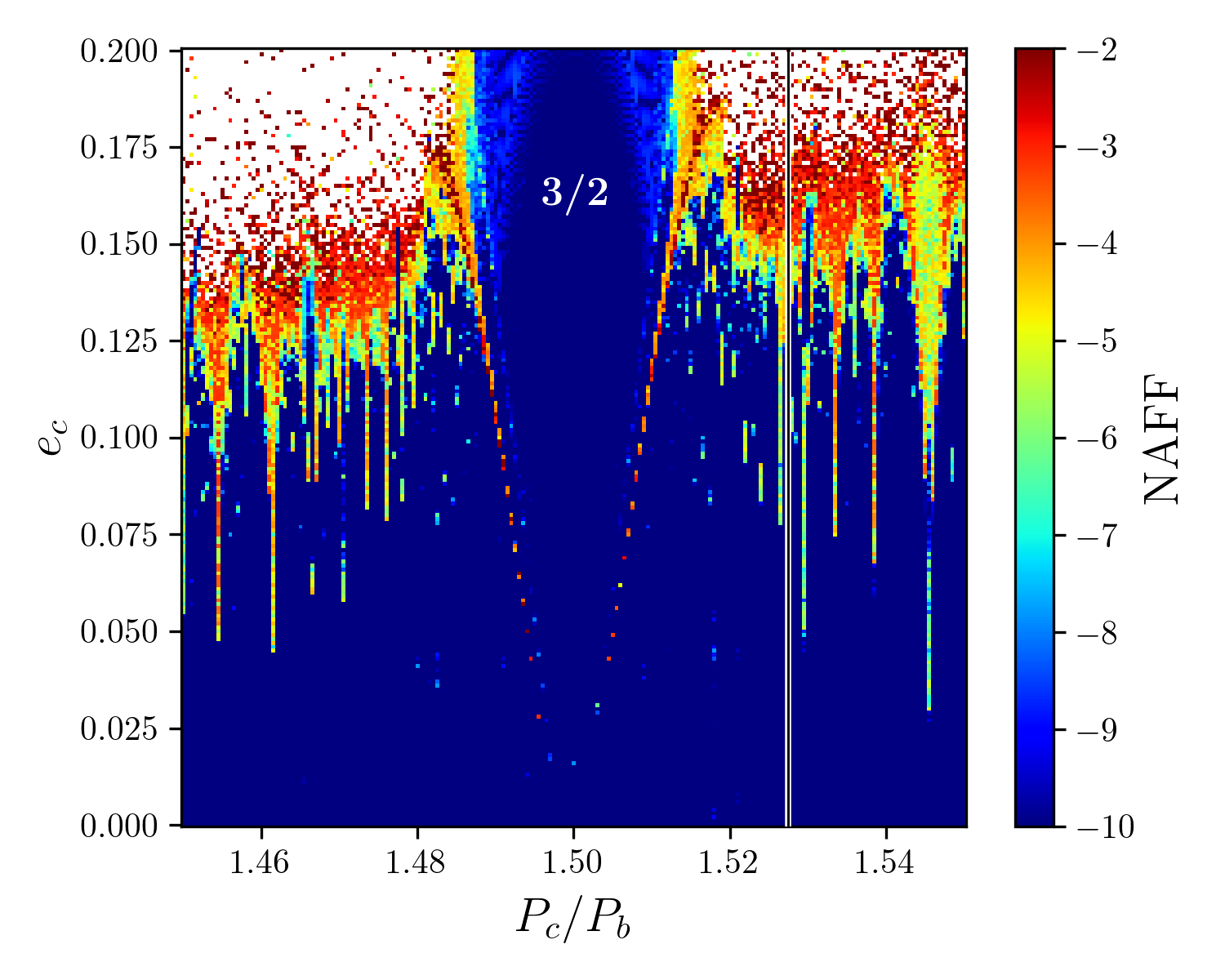}
\caption{Chaos map of the TOI-1453 system, with a resolution of 201$\times$201. This map estimates the chaos in a 2D section of the parameter space, along $P_c/P_b$ (x-axis) and $e_c$ (y-axis). The colour code depicts the NAFF chaos estimation, where redder areas are more chaotic. The white regions were unstable before the end of the short-term numerical simulations (escape or close encounter). The actual period ratio of the planet pair constrained from our photometric fit is represented by the vertical line.} 
\label{Fig:ChaosMap} 
\end{figure}

In Fig.\,\ref{Fig:ChaosMap}, we present the resulting grid of system configurations. The colour is indicative of NAFF, with redder regions denoting more chaotic systems. White areas indicate strongly unstable configurations, that did not reach the end of the numerical simulation (escape or close encounter as defined by the mutual Hill radius between planets b and c). Higher eccentricity orbits naturally lead to stronger dynamical chaos and potentially short-term instability. The larger the period ratio, the weaker the planet-planet gravitational interactions. As such, towards the right end of the map, the systems harbour a weakly chaotic behaviour for a wider range of initial orbital eccentricities. This is illustrated in Fig.\,\ref{Fig:ChaosMap} by the positive slope of the limit between blue and red, and red and white systems. 
The vertical line informs about the position of the system on the x-axis, which is well-constrained by the transit observations. At that position, we observe a rough upper limit of $e_c$<0.12 to keep a weak chaos. We stress however that this limit is approximate. While calibrating NAFF on the actual orbital stability is possible \citep{Couetdic2010, Stalport2022}, the purpose of chaos maps is to highlight the dynamical topology of the parameter space. The blue areas are feasible solutions, while the other more chaotic configurations will ultimately lead to orbital instability within short timescales. 
There exists rigorous dynamical techniques to refine the planet parameters based on reliable parameter explorations \citep[e.g.][]{Tamayo2020b, Stalport2022}, but this is beyond the scope of the present investigation. The 3:2 MMR appears in Fig. \ref{Fig:ChaosMap} as the vertical V-shape feature of stable systems extending to higher eccentricities. As this chaos map shows, the planets in TOI-1453 lie outside of the 3:2 MMR. 

\subsection{TTV investigations} \label{Sect:TTV}
In an attempt to further constrain the mass of TOI-1453 b, we used the large TESS dataset to explore potential TTVs of the outer planet TOI-1453 c. Due to the gravitational interactions between the two planets, their transit times should deviate from the pure Keplerian ephemerides with an amplitude that depends chiefly on the eccentricities, the mass ratios, and the orbital periods. While the transits of TOI-1453 b are too shallow for measuring their individual timings, we investigated potential TTVs on the transits of TOI-1453 c as a result of the planet-planet gravitational interactions. 

\begin{figure} 
    \centering
    \includegraphics[width=\columnwidth]{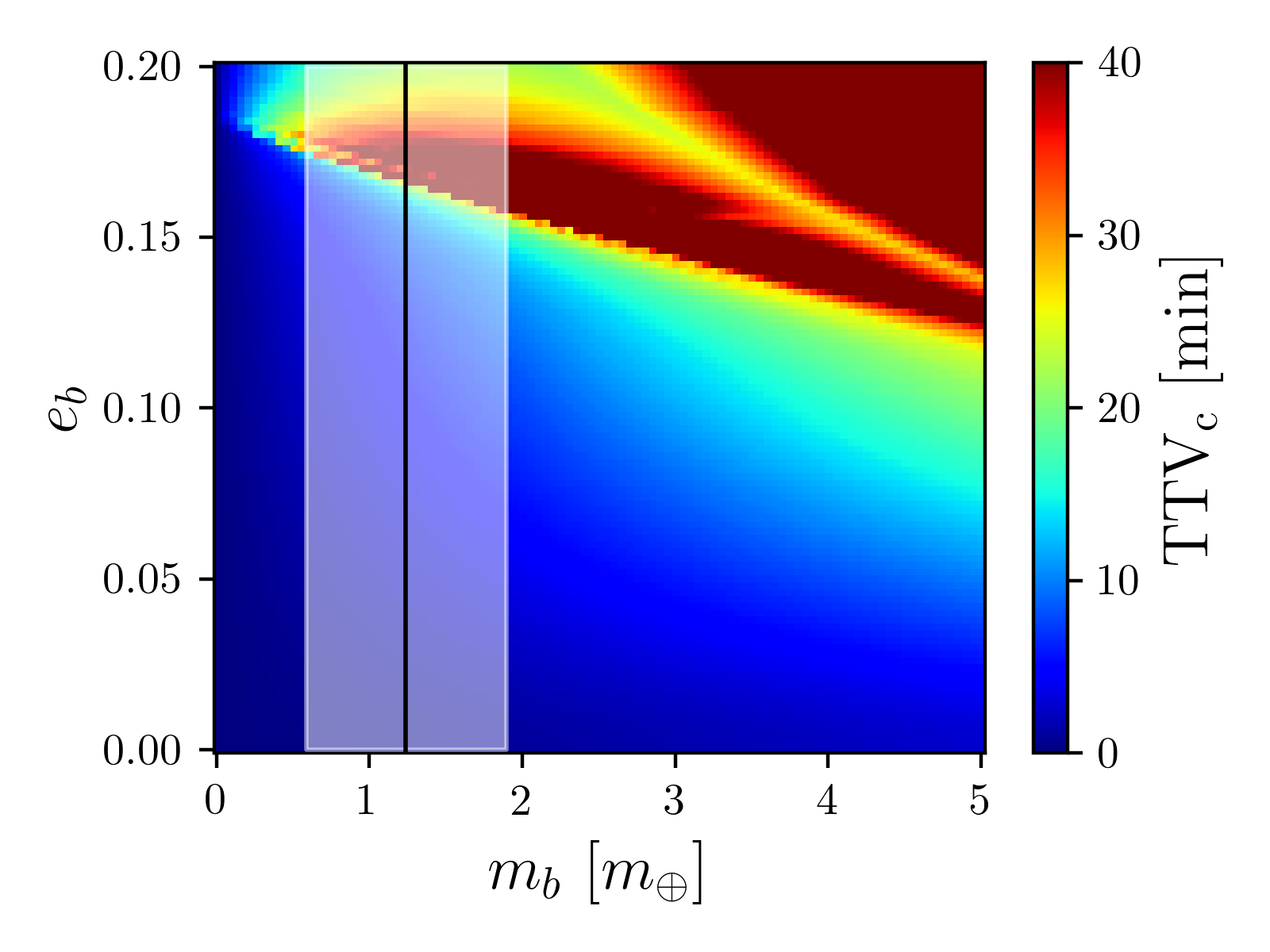}
\caption{Grid of simulated TTVs of TOI-1453 c as a function of the mass of the inner planet $M_b$ and its orbital eccentricity $e_b$. The vertical black line depicts the best-fit mass estimate from our joint fit, while the surrounding white band represents the 68.27$\%$ confidence interval on this estimation.} 
\label{Fig:TTVsimu} 
\end{figure}

The planets are close to, but outside of the 3:2\,MMR. Therefore, their TTV amplitudes should be modest. We carried out numerical simulations to estimate the expected TTV amplitudes in the TESS dataset. To proceed, we explored the influence of the mass and the orbital eccentricity of the inner planet on the TTVs of the outer one, via the design of a 101$\times$101 grid of simulations. For each combination of parameters, we used \texttt{TTVFast} \citep{Deck2014} to integrate the system on the timespan of our TESS observations (i.e., 1711 days) with a time step of 0.05 day. We verified the validity of this time step by performing the same integration with smaller time steps, and obtained similar results. During the integration, we recorded the transit times of TOI-1453\,c and fitted this time series with a line, the slope of which gives the orbital period of the linear ephemerides. We retained the maximum deviation among the transit times from this linear ephemerides, and repeated this process for each point in the grid. The resulting map of simulated TTV amplitudes is presented in Fig.\,\ref{Fig:TTVsimu}. While a large range of TTV amplitudes is observed throughout the grid, most of the parameter space is consistent with TTVs smaller than 20 minutes in amplitude. Notably, within the 68.27$\%$ range of masses $M_b$ derived from our joint fit, we observe small amplitude ($\lesssim$20 min) TTV up to an orbital eccentricity of $e_b \lesssim$\,0.15. 

\begin{figure} 
    \centering
    \includegraphics[width=\columnwidth]{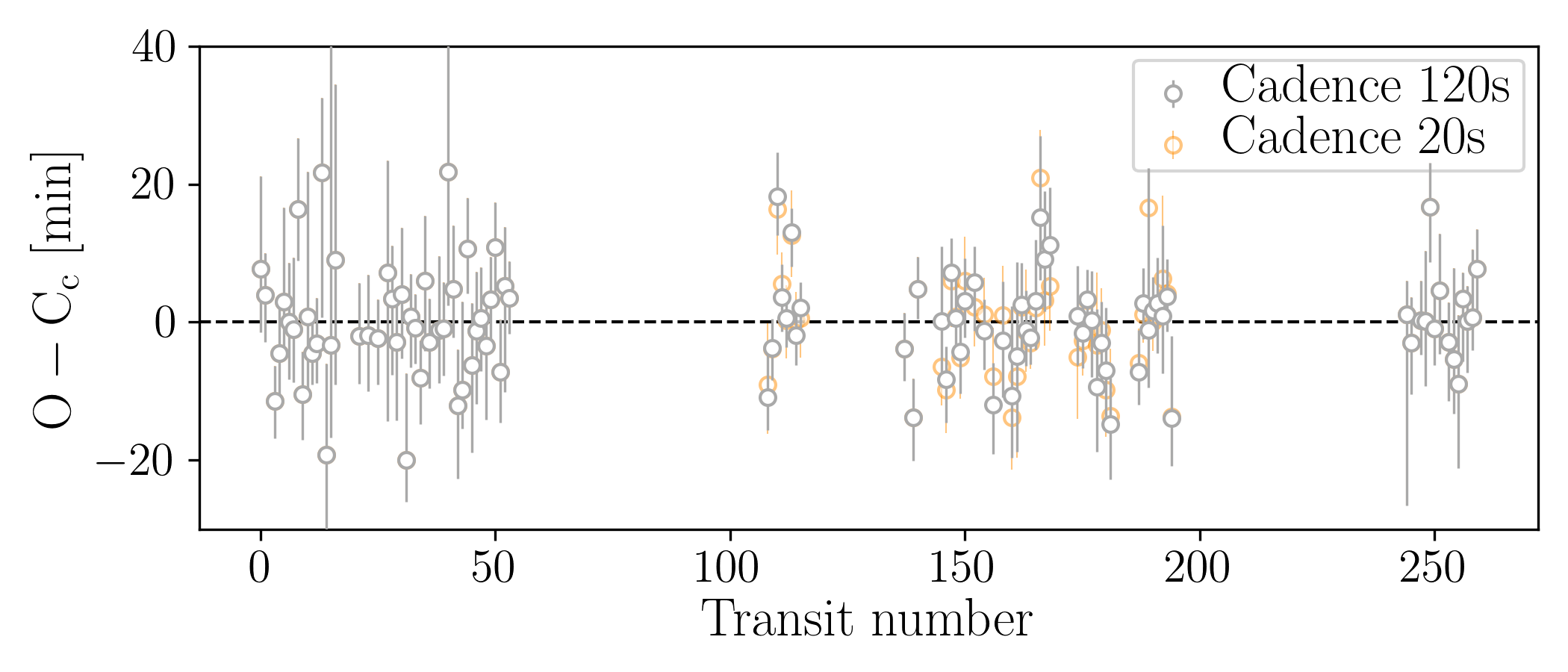}
    \includegraphics[width=\columnwidth]{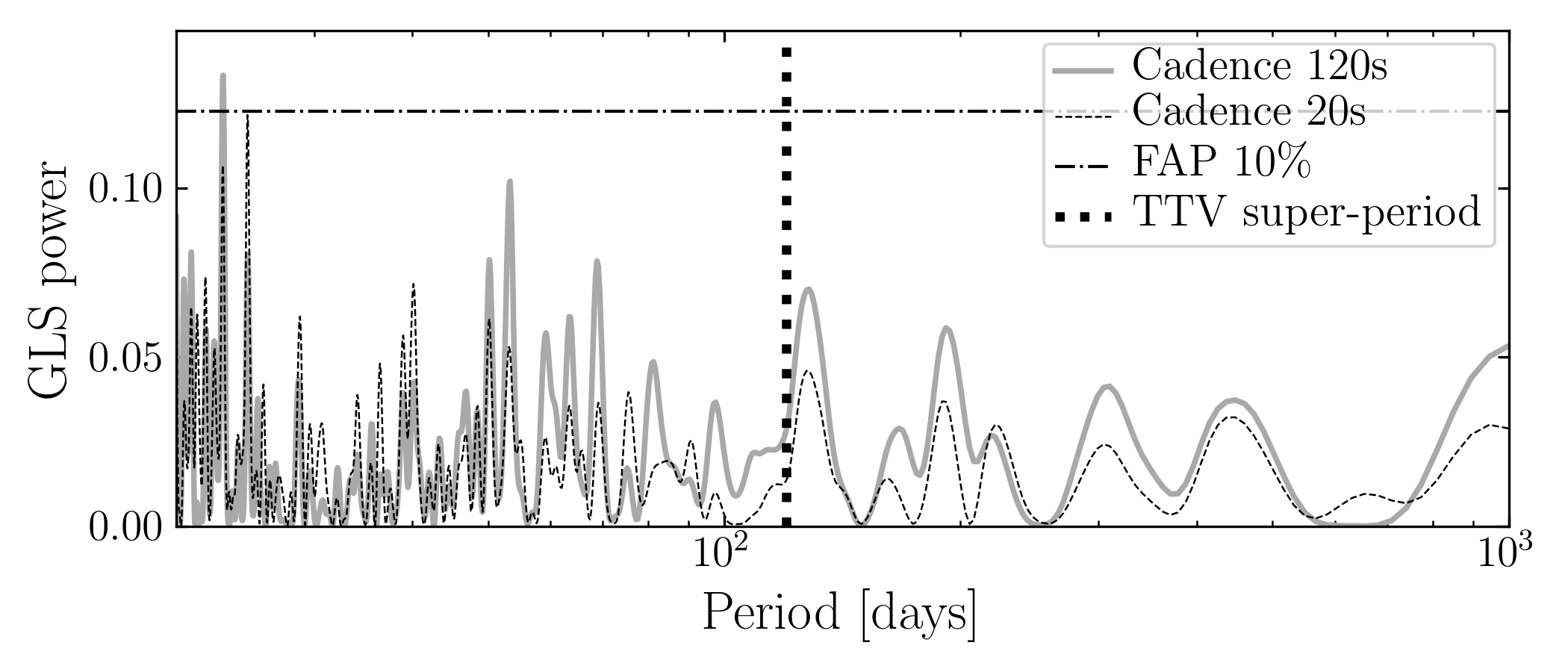}
    \includegraphics[width=\columnwidth]{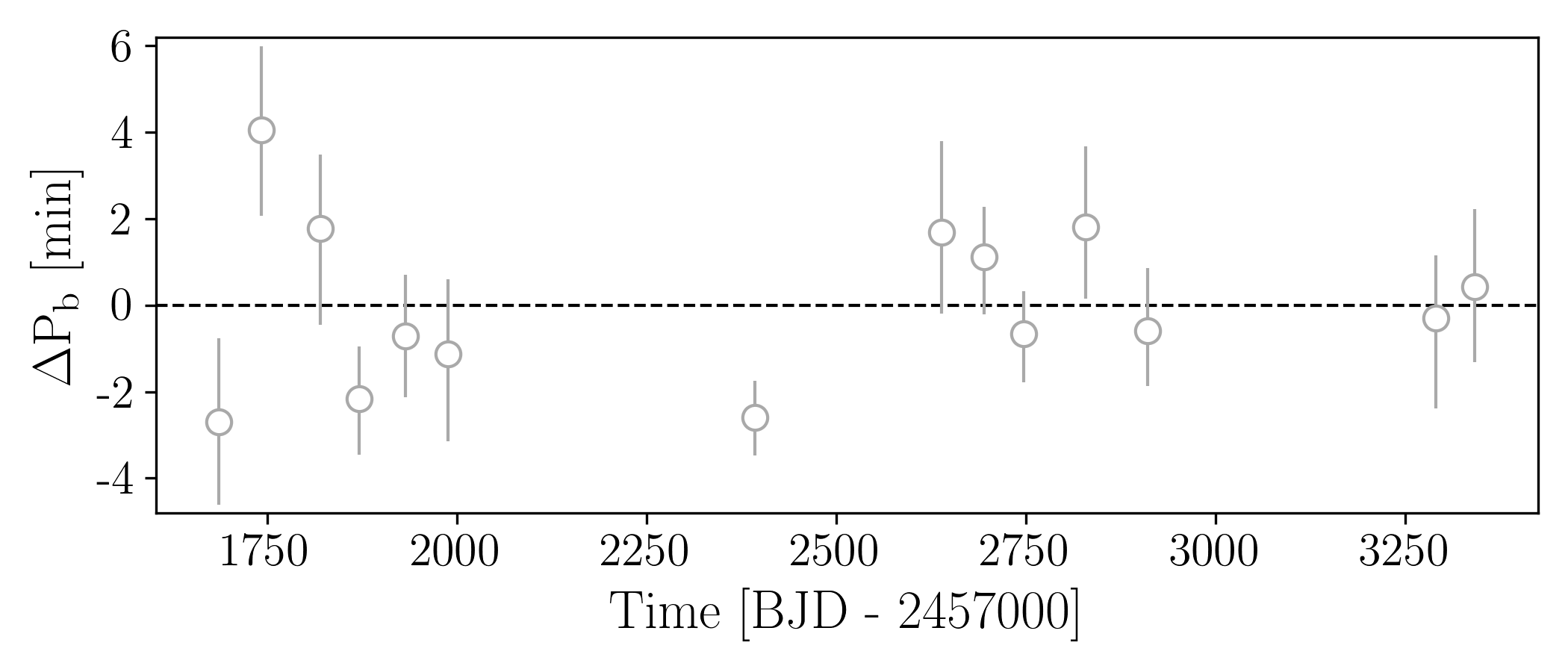}
\caption{TTV searches in the TESS LCs. The top plot shows the O-C diagram of the individual transits of planet c, and the middle panel presents the GLS periodogram of those observations. The solid line is the periodogram derived from all the 2min cadence data. The dashed line was obtained after replacing the 2min by 20s cadence data when available, and proceeding to new transit fits (cf. the orange points in the top plot). We do not find significant periodicity. The bottom plot presents the variation of the orbital period of planet b as measured from pairs of consecutive sectors (with 2min cadence).} 
\label{Fig:TTVobs} 
\end{figure}

With these prospects in mind, we investigated TTVs in the TESS data via a per-sector fit of the transit times. Our transit model includes TOI-1453\,b with linear ephemerides, while the outer planet has varying transit times. Indeed, instead of providing the parameters $P_c$ and $T_{0c}$, we added each transit time as a free parameter. Therefore, we included as many parameters as there are transits of planet c in the considered TESS sector. For both planets, we used the results of our joint global fit to impose narrow Gaussian priors on the $r1$ and $r2$ parameters, as well as $P_b$ and $T_{0,b}$. Our analyses were carried out with \texttt{Juliet} and the \texttt{dynesty} nested sampling algorithm. We undertook a two-step process to precisely constrain the transit times of TOI-1453\,c. For each sector, we performed a first fit with wide Gaussian priors on the individual transit times (0.1d of standard deviation), with the mean of the Gaussian derived from the linear ephemerides prediction. These results were used to narrow down the priors on the transit times for a second fit. We retained the results of that second fit to estimate the individual transit times. At the end of this process, we obtained the individual transit times of TOI-1453\,c in all the sectors. In Fig.\,\ref{Fig:TTVobs}, we present the deviation of each transit time with respect to the linear ephemeris, well-known as the O-C diagram (Observed timings minus the Calculated ones). It does not show large variations in the transit times: if any, their amplitudes are smaller than 20 minutes. This is in line with the results of our dynamical simulations.

Even if the amplitude of the TTV signal is expected to be small, we checked if the time series of O-C transit times could harbour a significant periodicity. We computed the GLS periodogram \citep[GLS; ][]{Zechmeister2009} of these transit times, and find the absence of significant periodicity as shown in the middle panel of Fig.\,\ref{Fig:TTVobs}. The highest peak slightly exceeds a False Alarm Probability (FAP) of 10$\%$ only. However, due to the proximity of the system to the 3:2 MMR, we note that one of the peaks lies close to the expected TTV super-period associated with this resonance. The TTV super-period is a long-term periodic variation of the TTV pattern often dominant in amplitude, and caused by the proximity of a pair of planets to a MMR. Close to a j:k MMR, the angles $j\lambda_c-k\lambda_b-(j-k)\varpi_b$ and $j\lambda_c-k\lambda_b-(j-k)\varpi_c$ vary slowly, where $\lambda$ and $\varpi$ depict the mean longitude and the longitude of periastron, respectively (in the co-planar case, the longitude of periastron is equivalent to the argument of periastron $\omega$). Consequently, the two planets regularly end up at the same relative position with respect to each other, which creates a constructive effect. Due to these strong mutual gravitational interactions, the planets will deviate significantly from their Keplerian orbits on a timescale given by the TTV super-period \citep[e.g.][]{Agol2018}:  
$$ P_{\rm TTV} ~ = ~ 1 / |j/P_c - k/P_b|. $$ 
The further away the planet pair from the resonance, the shorter in time and smaller in amplitude is this effect. In the case of TOI-1453, the planet pair has an expected super-period of 120 days due to its proximity to the 3:2 MMR. This is represented by the vertical dotted line in Fig.\,\ref{Fig:TTVobs}. 

To further investigate the tentative TTV periodicity close to the expected super-period, we also analysed the 20-second cadence TESS photometry when available. Our motivation was to attempt to reduce the uncertainties on the transit times and maybe increase the significance of the observed signals. However, the 20s TESS photometry is not available for all the sectors. Particularly, the data products are available in 20s cadence for the following sectors: 40, 41, 49, 50, 51, 52, 53, 54, 56, 57, 59, and 60 (12 sectors out of 29). For each of them, we repeated the process outlined above to retrieve precise transit times, and added these timings to the sectors that only contain 120s data. The new periodogram is presented in Fig.\,\ref{Fig:TTVobs} as a dashed line. It unambiguously shows that the significance of the signal decreases close to the TTV super-period. Additionally, we note that all the peaks are located below the FAP 10$\%$ line, making them particularly insignificant. In conclusion, with our current dataset we do not detect significant TTVs on TOI-1453 c. The use of advanced TTV detection and characterisation techniques, such as RIVERS \citep{Leleu2021}, could potentially unveil a TTV signal and constrain the mass of TOI-1453\,b. 

The mass of TOI-1453\,b is likely smaller than TOI-1453\,c, given its smaller radius and non-detection in the RV data. Therefore, the TTV amplitude on TOI-1453\,b due to planet-planet interactions should be larger than on TOI-1453\,c. As we already indicated, the individual transits of TOI-1453\,b are too shallow to constrain their TTVs. As a last investigation, we explored an alternative route by stacking the transit times of TOI-1453\,b in all the pairs of sectors, and searched for potential orbital period variations as a proxy for TTVs. To proceed, we performed linear ephemerides fits on all the pairs of consecutive sectors (sectors 14-15, 16-17, 19-20, 21-22, 23-24, 25-26, 40-41, 49-50, 51-52, 53-54, 56-57, 59-60, 73-74, 75-76) to derive a total of 14 period estimates. With datasets of two sectors, we can stack enough transits of the inner planet to obtain a robust period estimation. For each pair of sectors, we undertook a two-step fitting procedure again. First, we carried out a fit with a wide prior on $\rm P_b$ (normal prior with standard deviation of 0.1 day) and centred on the linear ephemerides expected value obtained from our global joint fit. Secondly, we used these results to set narrow priors on $\rm P_b$ for a new fit, and saved the results of this second fit. We present the time series of the period variations $\Delta\,P_b$ compared to the linear ephemerides in the bottom panel of Fig.\,\ref{Fig:TTVobs}. Throughout the TESS observations, we observe a maximum point-to-point period variation of 7 min. Compared to the linear ephemerides of our global fit, we do not observe period deviations larger than 4 min. Furthermore, there is no visible correlation between these small variations and the TTV diagram of the outer planet. Hence again, we do not find significant TTV signal. The absence of significant orbital period variations also highlights the robustness of our planet radius estimations. The TTV, if any, are not large enough to blur the signal of stacked transits.

%%%%%%%%%%%%%%%%%%%%%%%%%%%%%%%%% Jo Ann analyses
\section{Interior modelling} \label{Sect:Interior}
\begin{figure*}
    \centering
    \includegraphics[width=\linewidth]{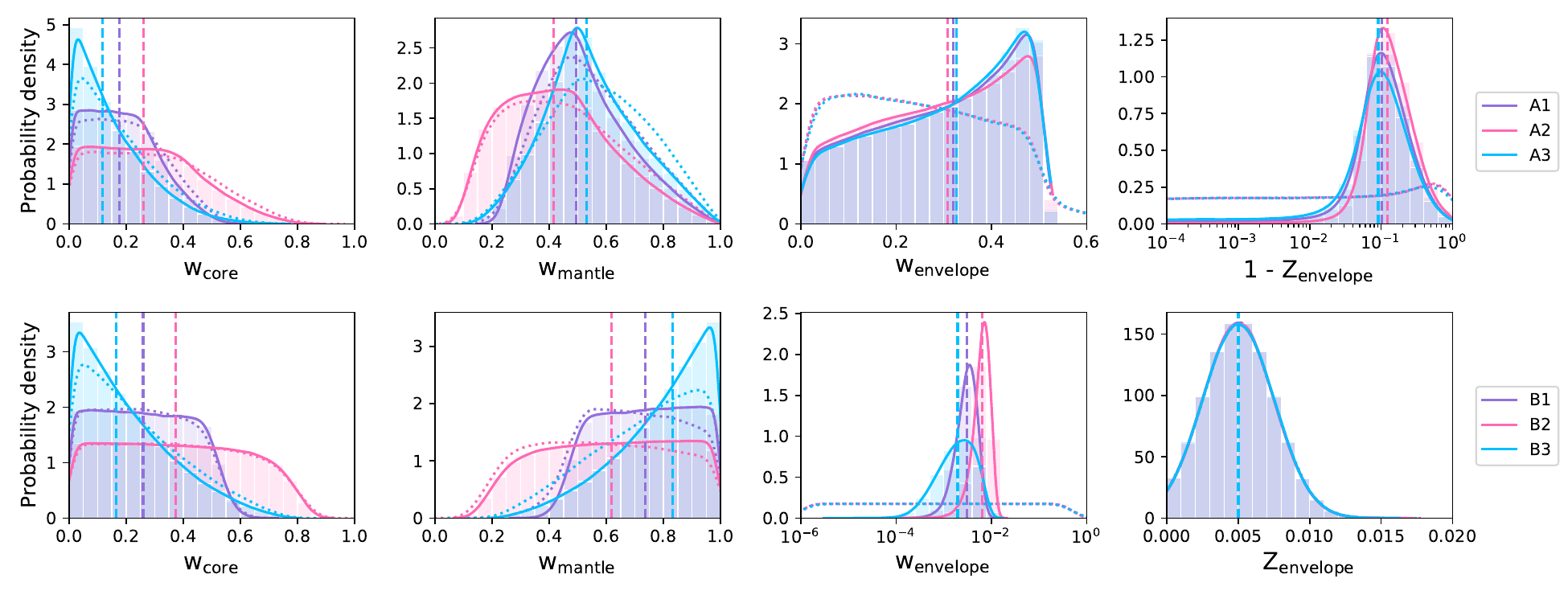}
    \caption{Posterior distributions of the interior structure of TOI-1453 c, showing mass fractions of the inner core, mantle, and envelope, as well as the water fraction in the envelope. The top row corresponds to a water prior compatible with a formation scenario beyond the ice line, while the bottom row assumes a water-poor composition, in agreement with formation inside the ice line. Colours represent three priors for the planetary Si/Mg/Fe ratios: stellar (purple, A1 and B1), iron-enriched (pink, A2 and B2), and uniformly sampled from a simplex (blue, A3 and B3). Dashed lines mark the median of each distribution, with priors indicated by dotted lines.}
    \label{fig:int_struct_c}
\end{figure*}

We applied the publicly available Bayesian modelling framework \texttt{plaNETic}\footnote{\url{https://github.com/joannegger/plaNETic}} \citep{Egger+2024} to model the interior structure of TOI-1453~c. In this framework, the planet is modelled as a combination of an inner iron core with up to 19\% of sulphur, a silicate mantle made up of SiO$_2$, MgO and FeO, and a volatile layer made up of a uniform mixture of H/He and water. The \texttt{plaNETic} code is split into an inverse model that samples possible internal structure parameters using the chosen priors, and a forward model, that calculates the transit depth of each sampled structure, which is then compared to the observed value. This process is sped up by replacing the forward model (the planetary structure model of the BICEPS code, \citealp{Haldemann+2024}) with a neural network, which allows for a fast and reliable characterisation of the planetary interior.

We ran six different models assuming different priors. First, we vary the priors for the planetary Si/Mg/Fe values, assuming they are stellar \citep[e.g.][]{Thiabaud+2015}, iron-enriched compared to the host star \citep[e.g.][]{Adibekyan+2021}, or leaving them unconstrained by sampling uniformly from a simplex with an upper limit of 0.75 for the iron fraction. On the other hand, we also vary the priors for the water content in the planet, once assuming that water was readily available for accretion during the planets formation, and once that water could only be accreted through the accreted gas, compatible with a formation inside the water iceline.

Figure~\ref{fig:int_struct_c} shows the resulting posterior distributions of the most important interior structure parameters, while a summary of the results for the full set of parameters can be found in Table~\ref{tab:int_struct_c}. We find that, for both chosen water priors, we do not constrain the mass fractions of the inner core and mantle layers, for which the posteriors largely agree with the priors. For the water-rich prior (top row), we find that a wide range of envelope mass fractions is compatible with the observed planetary properties, with a preference for higher envelope mass fractions. Most of these inferred envelopes are water-rich, with a median envelope water mass fraction of around 90\%. Conversely, for the water-poor prior (bottom row), we inferred tightly constrained metal-poor envelopes with mass fractions of $0.3_{-0.1}^{+0.2}$\%, $0.6_{-0.2}^{+0.3}$\%, and $0.2_{-0.1}^{+0.3}$\%, for the stellar, iron-enriched and free Si/Mg/Fe priors respectively. 

Finally, we also applied \texttt{plaNETic} to model TOI-1453\,b. However, we could not set any constraints on its structure due to the large uncertainties onto the planet mass.

\section{Discussion and conclusion} \label{Sect:Conclusion}
In this paper, we validated two planets orbiting TOI-1453 based on TESS observations, ground-based photometric, spectroscopic, and high-resolution imaging follow-up. We then confirmed the outer planet with 100 HARPS-N spectra. We further analysed the 29 TESS sectors and the HARPS-N data to characterise this system. We summarise our findings below. 

\begin{figure*}[t!] 
    \centering
    \includegraphics[width=0.8\columnwidth]{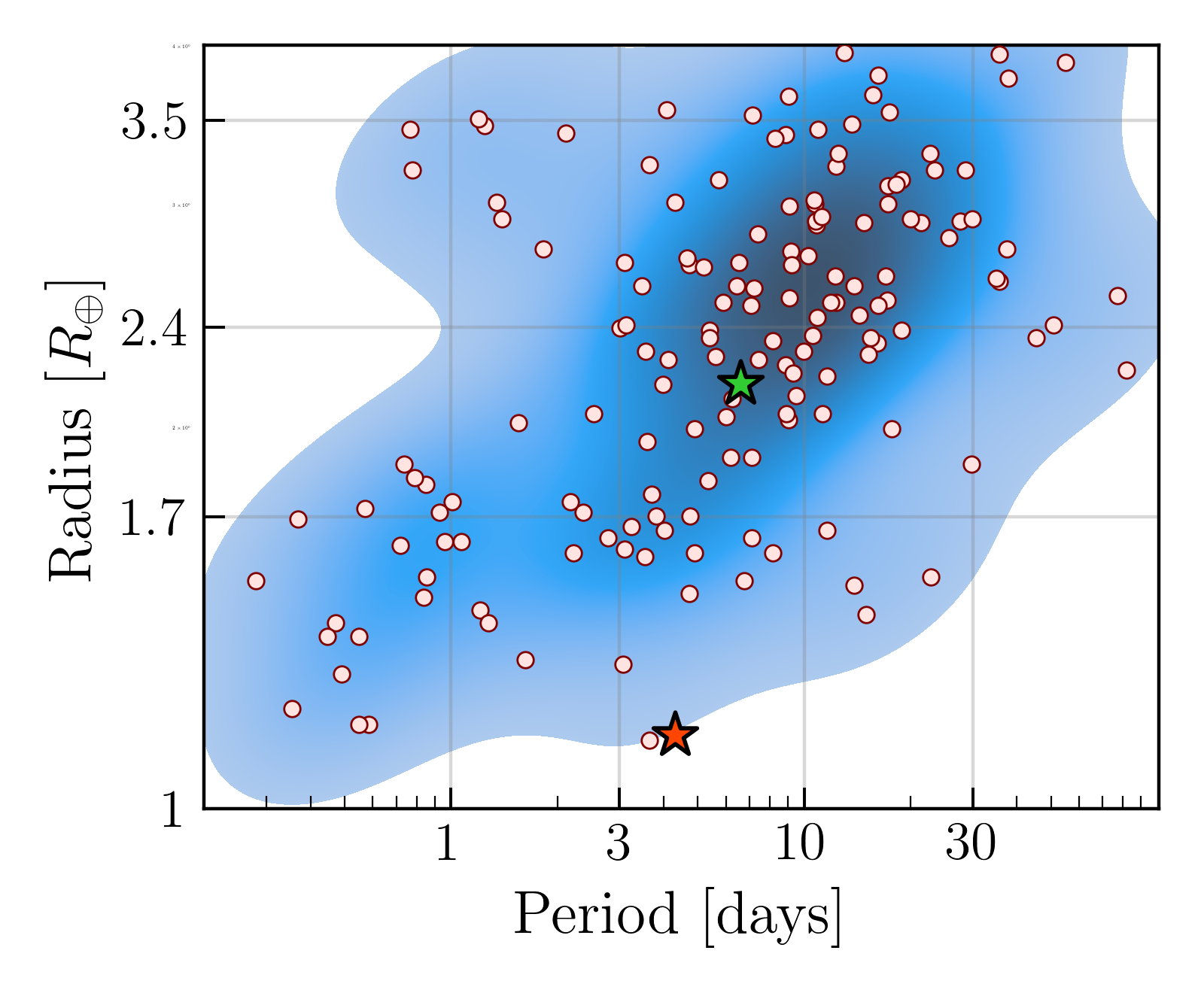}
    \includegraphics[width=0.8\columnwidth]{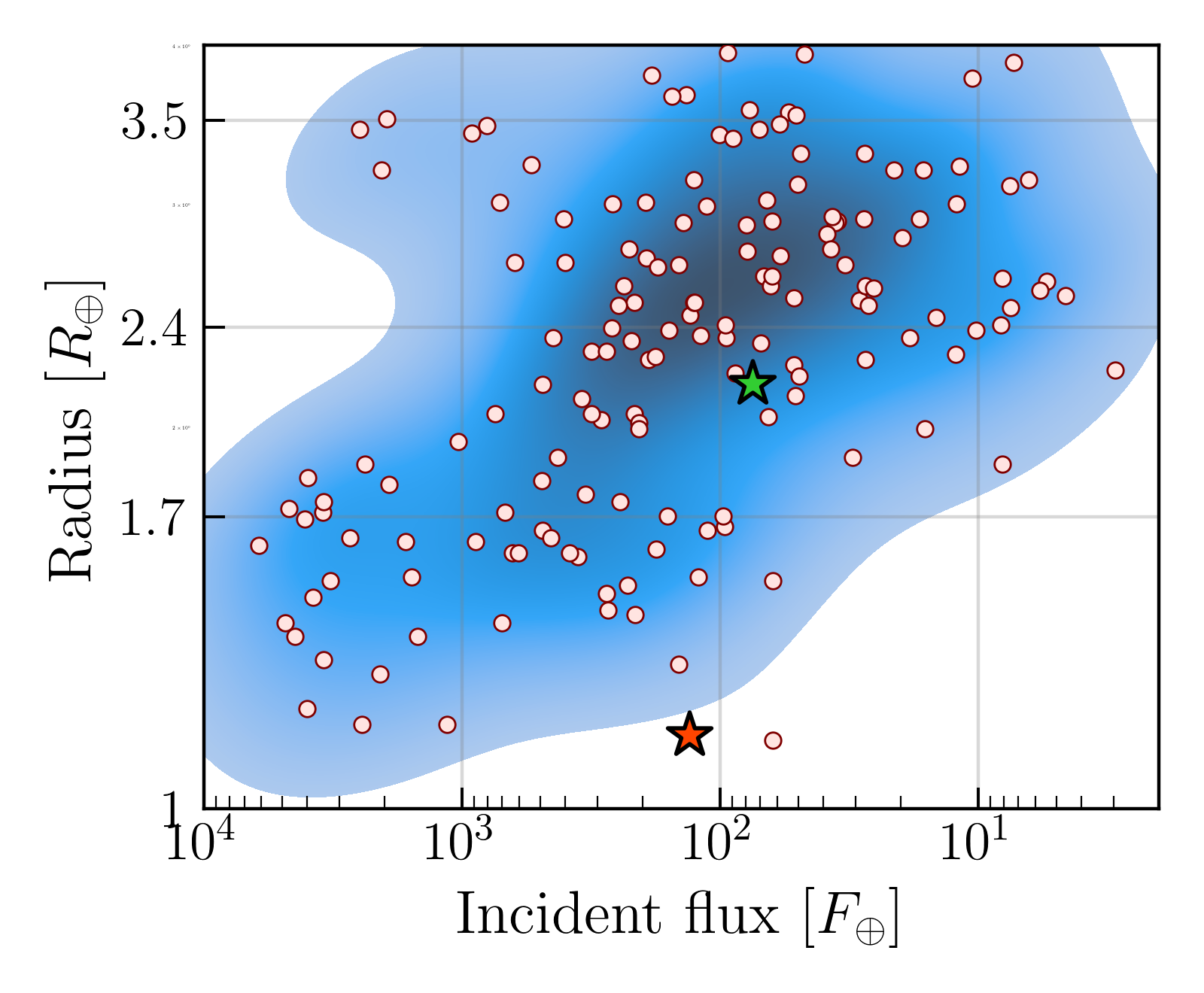}
\caption{Population of precisely characterised planets (uncertainty on density smaller than 25$\%$) orbiting FGK stars. The left panel presents the period-radius diagram, while the right panel shows the incident flux versus radius diagram. The shaded areas indicate the planet occurrence rate. The orange (green) star symbol represents TOI-1453 b (c) in these 2D spaces. } 
\label{Fig:RadValley} 
\end{figure*}

\begin{itemize}
\item TOI-1453 is a metal-poor K-dwarf star with a mass and radius of 0.715$\pm$0.035\,$M_{\odot}$ and 0.720$\pm$0.029\,$R_{\odot}$, respectively. It belongs to the Galactic thin disc at a distance of $\sim$78.9\,pc. This star is orbited by a stellar companion with a current angular separation of 1.9$''$. This corresponds to a projected distance of $\sim$150\,AU. \citet{Christian2024} found that the binary orbit is eccentric and co-planar with the planetary orbits and constrained the semi-major axis of this binary to 171$\substack{+127 \\ -58}$AU. TOI-1453 appears very inactive in the time span of our observations. 
\item TOI-1453\,b orbits the host star with an orbital period of $P_b$=4.314\,days. It is a super-Earth with estimated mass and radius of $M_b$=1.24$\substack{+0.66 \\ -0.65}$\,$M_{\oplus}$ and $R_b$=1.173$\substack{+0.059 \\ -0.056}$\,R$_{\oplus}$, respectively. Its bulk composition is poorly constrained given the currently low precision on the planet mass (1.9$\sigma$ detection). However, it is likely rocky with a bulk density of $\rho_b$=0.76$\substack{+0.44 \\ -0.40}$\,$\rho_{\oplus}$.  
\item TOI-1453\,c orbits the host star with an orbital period of $P_c$=6.589\,days. It is a sub-Neptune with estimated mass and radius of $M_c$=2.95$\substack{+0.83 \\ -0.84}$\,M$_{\oplus}$ and $R_c$=2.224$\pm$0.095\,R$_{\oplus}$, respectively. It harbours a very low bulk density ($\rho_c$=0.267$\substack{+0.087 \\ -0.080}$\,$\rho_{\oplus}$), compatible with either a water-rich composition or the presence of an extended $H$ atmosphere on top of a rocky planet. TOI-1453 c is one of the rare sub-Neptunes below 3M$_{\oplus}$ known to date. 
\end{itemize}

TOI-1453 b and c span the radius valley, as illustrated in Fig. \ref{Fig:RadValley}. These plots present the current sample of precisely characterised planets orbiting FGK stars (\teff$>$4000 K), obtained with the DACE platform\footnote{\url{https://dace.unige.ch}}. Projected along the period (left) or incident flux (right) versus radius spaces, the planet population reveals a characteristic occurrence gap between 1.7 and 2 R$_{\oplus}$ \citep{Fulton2017, VanEylen2018, Fulton2018}. TOI-1453 b, represented by the orange star symbol, clearly stands below the gap. While TOI-1453 c (the green star symbol) stands above the radius valley close to the edge. It is noteworthy to observe that there are very few planets with radius $\sim$1R$_{\oplus}$, and TOI-1453 b occupies a yet-empty area. This is likely the result of the low S/N that these planets have around FGK stars; whereas significantly more detections were made around M dwarfs \citep[e.g.][]{Kemmer2022, Gillon2024}. 
 Constraining and understanding the composition of TOI-1453\,c and the possible contrast with TOI-1453\,b further are of particular importance. Currently, there are two interpretations for the population of sub-Neptunes. On one side, they could be 'dry', water-poor, puffy worlds. In this scenario, the super-Earths were formerly sub-Neptunes that lost their atmospheres due either to photo-evaporation \citep{Owen2013} or core-powered mass loss \citep{Ginzburg2018}. 
On the other hand, sub-Neptunes could be water-rich worlds, as a result of formation beyond the iceline followed by inward migration \citep[e.g.][]{Luque2022}. 

The proximity to MMR is indicative of migration in the protoplanetary disk. \citet{Chen2024} identified in the Kepler sample a larger proportion of sub-Neptunes close to a MMR compared to super-Earths. They use this observation and planetary formation models to propose that both ex situ and in situ formation mechanisms shaped the sub-Neptunes population (some formed beyond the ice line, others inside, following distinct formation routes). As was highlighted in \citet{Leleu2024}, close-to-MMR sub-Neptunes are less dense. The bulk density of TOI-1453 c supports this trend, yet further detailed investigations are needed, such as atmospheric characterisation, to lift the compositional degeneracy. Additionally, stellar metallicity is expected to impact the planet compositions \citep[e.g.][]{Santos2017}, with metal-poor stars expected to harbour planets with small core-mass fractions. While the internal structure modelling was unable to constrain the core mass fraction, the low metallicity estimated in this work could partially contribute to the small bulk density of TOI-1453 c. 

The fact that TOI-1453 b and c are close to but still beyond a MMR is likely a consequence of tidal dissipation. At small orbital distances, tidal torques can move planet pairs away from the resonance by increasing their period ratio \citep[e.g.][]{Delisle2014a, Louden2023}. The Kepler sample supports this tidal MMR breaking. Indeed, \citet{Delisle2014} observe a net preference for MMR offsets in short-period planet pairs as opposed to pairs with P$>$15d. The planetary tides are expected to have synchronised the planet spins with their orbital frequencies to reach a state of tidal locking \citep[e.g.][]{Bolmont2014}. However, the circularisation timescale can be significantly larger. There is no hint of orbital eccentricity in the TESS and HARPS-N observations. According to our dynamical analysis, any potential eccentricity should not exceed 0.12 to enable an orbital stability (cf. Fig. \ref{Fig:ChaosMap}). Small orbital eccentricities are also expected from tidal dissipation, given the small orbital separation of the planets. This tidal eccentricity damping counteracts the excitation from planet-planet gravitational interactions, leading to a small but non-zero equilibrium value \citep{Mardling2007}. Using the analytical estimation of \citet{Matsumura2008}, we estimated the timescale to reach tidal circularisation. Assuming an initial small orbital eccentricity of 0.01, a constant tidal quality factor $Q$ of $10^4$ \citep{Louden2023} and a Love number identical to the Earth ($k_2$=0.3), we obtain $\tau_{\rm circ,c} \sim 5-6\times10^9$ yr for TOI-1453 c. Regarding TOI-1453 b, the same exercise leads to $\tau_{\rm circ,b} \sim 9-10\times10^9$ yr. While these timescales are approximative, they suggest that a small orbital eccentricity could be maintained. 

Furthermore, the nearby stellar companion is likely to have played a role in shaping the architecture of the system we observe today. To some extent, it could have had a long-term impact on the planetary dynamics, such as inducing a spin-orbit misalignment or low-frequency variations of the planetary orbital elements. Recent population studies revealed an alignment of binary orbits with the orbits of small S-type planets \citep[e.g.][]{Behmard2022, Christian2024} for binary separations comprised between 100 and 700\,AU. This result supports theoretical investigations predicting that the binary companions can align the protoplanetary disc inside which the planets form \citep{Batygin2012}. But also, the gravitational influence of this companion could have truncated the protoplanetary disk around TOI-1453 and subsequently influenced the formation of the planets \citep[e.g.][]{Ronco2021}. This exact influence is yet poorly understood, as systematic studies on S-type planets (which orbit one component of a multi-star system) are just beginning. \citet{Sullivan2023} analysed the population of S-type planets from the Kepler mission and found no radius valley. However, the number of precisely characterised S-type planets is yet very small. New detections, such as that of TOI-1453, are crucial to increase the sample size and firmly answer  questions related to the radius valley in S-type planets. 

\begin{figure} 
    \centering
    \includegraphics[width=\columnwidth]{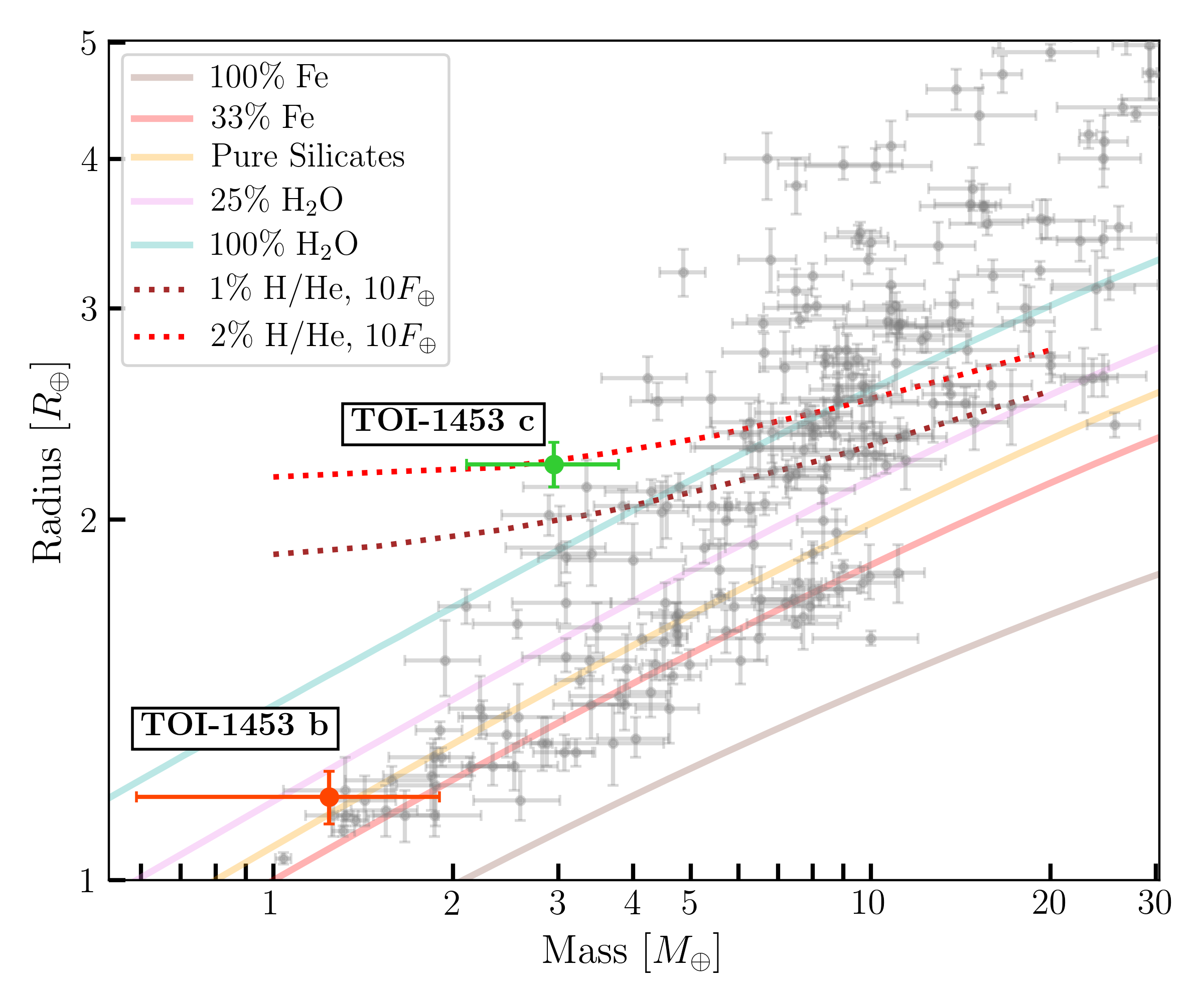}
\caption{Mass-radius diagram of all exoplanets smaller than 5$R_{\oplus}$ and with a precision on the mass and radius better than 25$\%$ and 8$\%$, respectively.} 
\label{Fig:MRdiagram} 
\end{figure}

In Fig. \ref{Fig:MRdiagram}, we present the current mass-radius diagram of small planets with a precision on the mass and radius better than 25$\%$ and 8$\%$, respectively \citep[cf.][]{Otegi2020, Parc2024}, and obtained via the DACE platform\footnote{\url{https://dace.unige.ch}}. We overplot in this diagram TOI-1453 b (in orange) and c (in green). Model compositional curves from \citet{Zeng2016} are also highlighted in the diagram with thick solid lines. TOI-1453\,b has likely a rocky composition as shown in the figure, even though this observation must be taken with caution given the large uncertainties on the planet mass. The dashed lines represent the mass-radius relations for Earth-like objects (33$\%$ Iron core plus 67$\%$ silicates mantle) with H/He atmospheres of different mass fractions. These relations, taken from \citet{Lopez2014}, take into account the thermal evolution of the planetary atmosphere, which is an important factor to interpret the mass and radius of sub-Neptunes \citep{Rogers2023}. In Fig.\,\ref{Fig:MRdiagram}, they are given for an irradiation of 10\,$F_{\oplus}$, a Solar metallicity and an age of 10\,Gyr. Compared to these curves, the mass and radius of TOI-1453\,c are compatible with an extended $H_2$ atmosphere. Naturally, these dashed lines are dependent on the stellar age, metallicity, incoming irradiation, and they do not take into account the photo-evaporation effect. Nevertheless, they illustrate the degeneracy in the mass-radius diagram between rocky cores with extended H/He atmospheres and water worlds. The distinction between these two compositional possibilities cannot be resolved from the mass and radius alone. Our modelling of the interior structure of TOI-1453\,c stressed this degeneracy (cf. Sect. \ref{Sect:Interior}). However, it allowed us to constrain the water mass fraction in the envelope both for a water-rich and a water-poor formation scenario. If planet c formed beyond the ice-line, we would expect a water content of $\sim$90$\%$. On the contrary, in a water-poor scenario we would expect an envelope mass fraction of $\sim$0.5$\%$, among which water would make up $\sim$0.5$\%$ only. 

An investigation of the atmosphere of the planet would provide important information in favour of one scenario or the other. To quantify the detectability of the atmospheres of these two planets, we computed the transmission spectroscopy metric \citep[TSM,][]{Kempton2018}, which is correlated with the expected S/N of the transmission spectroscopy signals. While the TSM is negligible for TOI-1453\,b given its small transit depth (TSM=5.1), we find a TSM of 78.8 for TOI-1453\,c. This is below the threshold of 90 recommened by \citet{Kempton2018} for planets of this radius range. However, we stress that this threshold serves as an indication only. Observations with JWST could provide high quality spectra of TOI-1453 and shed some light on the atmospheric composition of planet c. In Table \ref{tab:PlanetAtmParam}, we report additional planet properties useful for atmospheric follow-up. 

\begin{table}
\centering
\begin{threeparttable}
\caption{Additional planet parameters.}
\label{tab:PlanetAtmParam}
\begin{tabular}{@{}lcccp{6cm}@{}}
\toprule
Planet  &   Incident flux [$F_{\oplus}$]  &   $T_{eq}$ [$K$]    &   TSM    \\ \midrule 
TOI-1453 b & 132.6$\pm$10.4 & 944.2$\pm$17.8 & 5.1$\pm$0.1 \vspace{0.07cm} \\ 
TOI-1453 c & 75.3$\pm$5.8 & 819.8$\pm$15.6 & 78.8$\pm$1.5  \\ \bottomrule
\bottomrule
\end{tabular}
\end{threeparttable}
\end{table}

All these different characteristics (planet bulk densities, stellar host parameters, dynamics, stellar companion, atmospheric prospects, etc) constitute as many constraints that could be added to the various formation and evolution scenarios. They make TOI-1453 a particularly valuable system which, with its long photometric baseline, likely represents a PLATO archetype. 
PLAnetary Transits and Oscillations of stars (PLATO) is the forthcoming ESA space-based mission that will photometrically survey the same portion of sky for at least two consecutive years, with exquisite precision \citep{Rauer2014}. Thanks to this continuous monitoring, it will accumulate many transits of the short-period small-planet systems, similarly to TOI-1453. The extreme precision of PLATO will constrain further those systems, notably with a good estimate of the stellar age and rotation cycles, and very precise transit timings for TTV observations. The PLATO telescope will reveal more small-planet systems similar to TOI-1453 and amenable to RV follow-ups. This will populate the still sparse mass-radius region of small-mass sub-Neptunes, offering insights into the origin of these intriguing worlds.

\begin{acknowledgements}  
The authors thank the referee for a constructive report. 
This work is based on observations made with the Italian Telescopio Nazionale Galileo (TNG) operated by the Fundación Galileo Galilei (FGG) of the Istituto Nazionale di Astrofisica (INAF) at the Observatorio del Roque de los Muchachos (La Palma, Canary Islands, Spain). The HARPS-N project was funded by the Prodex Program of the Swiss Space Office (SSO), the Harvard University Origin of Life Initiative (HUOLI), the Scottish Universities Physics Alliance (SUPA), the University of Geneva, the Smithsonian Astrophysical Observatory (SAO), the Italian National Astrophysical Institute (INAF), University of St. Andrews, Queen’s University Belfast, and University of Edinburgh. 
This paper made use of data collected by the TESS mission and are publicly available from the Mikulski Archive for Space Telescopes (MAST) operated by the Space Telescope Science Institute (STScI). Funding for the TESS mission is provided by NASA’s Science Mission Directorate. We acknowledge the use of public TESS data from pipelines at the TESS Science Office and at the TESS Science Processing Operations Center. Resources supporting this work were provided by the NASA High-End Computing (HEC) Program through the NASA Advanced Supercomputing (NAS) Division at Ames Research Center for the production of the SPOC data products. 
This publication makes use of The Data $\&$ Analysis Center for Exoplanets (DACE), which is a facility based at the University of Geneva (CH) dedicated to extrasolar planets data visualisation, exchange and analysis. DACE is a platform of the Swiss National Centre of Competence in Research (NCCR) PlanetS, federating the Swiss expertise in Exoplanet research. The DACE platform is available at \url{https://dace.unige.ch}. 
M.S. thanks the Belgian Federal Science Policy Office (BELSPO) for the provision of financial support in the framework of the PRODEX Programme of the European Space Agency (ESA) under contract number C4000140754. 
A.M. acknowledges funding from a UKRI Future Leader Fellowship, grant number MR/X033244/1 and a UK Science and Technology Facilities Council (STFC) small grant ST/Y002334/1. 
M.C acknowledges the SNSF support under grant P500PT$\_$211024. 
KAC and CNW acknowledge support from the TESS mission via subaward s3449 from MIT. 
This research has made use of the Exoplanet Follow-up Observation Program (ExoFOP; DOI: 10.26134/ExoFOP5) website, which is operated by the California Institute of Technology, under contract with the National Aeronautics and Space Administration under the Exoplanet Exploration Program.
X.D acknowledges the support from the European Research Council (ERC) under the European Union’s Horizon 2020 research and innovation programme (grant agreement SCORE No 851555) and from the Swiss National Science Foundation under the grant SPECTRE (No 200021$\_$215200).
This work has been carried out within the framework of the NCCR PlanetS supported by the Swiss National Science Foundation under grants 51NF40$\_$182901 and 51NF40$\_$205606.
A.L. acknowledges support of the Swiss National Science Foundation under grant number TMSGI2$\_$211697. 
This work makes use of observations from the LCOGT network. Part of the LCOGT telescope time was granted by NOIRLab through the Mid-Scale Innovations Program (MSIP). MSIP is funded by NSF.
This article is based on observations made with the MuSCAT2 instrument, developed by ABC, at Telescopio Carlos S\'{a}nchez operated on the island of Tenerife by the IAC in the Spanish Observatorio del Teide. This work is partly financed by the Spanish Ministry of Economics and Competitiveness through grants PGC2018-098153-B-C31. As well, this work is supported by JSPS KAKENHI Grant Number JP24H00017, JP24K00689 and JSPS Bilateral Program Number JPJSBP120249910.
This work is partly supported by JSPS KAKENHI Grant Numbers JP17H04574, JP18H01265, and JP18H05439, Grant-in-Aid for JSPS Fellows Grant Number JP20J21872, JST PRESTO Grant Number JPMJPR1775, and a University Research Support Grant from the National Astronomical Observatory of Japan (NAOJ). This paper is based on observations made with the MuSCAT2 instrument, developed by ABC, at Telescopio Carlos Sánchez operated on the island of Tenerife by the IAC in the Spanish Observatorio del Teide.
SNQ acknowledges support from the TESS mission via subaward s3449 from MIT. 
This work has been partially supported by the National Aeronautics and Space Administration under grant No. NNX17AB59G, issued through the Exoplanets Research Program. 
This work has made use of data from the European Space Agency (ESA) mission {\it Gaia} (\url{https://www.cosmos.esa.int/gaia}), processed by the {\it Gaia} Data Processing and Analysis Consortium (DPAC,
\url{https://www.cosmos.esa.int/web/gaia/dpac/consortium}). Funding for the DPAC has been provided by national institutions, in particular the institutions participating in the {\it Gaia} Multilateral Agreement. 
M.P. acknowledge support from the European Union – NextGenerationEU (PRIN MUR 2022 20229R43BH) and the "Programma di Ricerca Fondamentale INAF 2023". 
\end{acknowledgements}

\bibliographystyle{aa} % style aa.bst
\bibliography{bib.bib}

\begin{appendix}
\onecolumn
\section{TESS and HARPS-N data}
In this appendix, we share a detailed summary of the TESS (Table \ref{tab:TESS_sectors}) and HARPS-N (Table \ref{tab:RV}) data used in this paper. The light curve of each TESS sector was obtained via a custom extraction from the TPF. Regarding the HARPS-N spectroscopic data, we report the YARARA dataset which we used in this work. The full table is available on the CDS. 

\begin{table*}[h!]
\centering
\begin{threeparttable}
\caption{Our custom extraction of the 29 TESS sectors analysed in this work.}
\label{tab:TESS_sectors}
\begin{tabular}{@{}lllllccp{6cm}@{}}
\toprule
Sector & Start date [UTC] & End date [UTC] & Nmeas & MAD [ppm]  & Additional transits b\tnote{1} & Additional transits c\tnote{1}    \\ \midrule \\
14  & 2019-07-18  & 2019-08-14 & 18417 & 707.8 &  0 & 0  \vspace{0.2cm}  \\
15  & 2019-08-15  & 2019-09-10 & 17740 & 707.5 &  0 & 1  \vspace{0.2cm} \\ 
16  &  2019-09-12 & 2019-10-06 & 16706 & 710.8 & 2 & 2  \vspace{0.2cm} \\ 
17  & 2019-10-08 & 2019-11-02 & 16639 & 714.7 & 2 & 2  \vspace{0.2cm}  \\ 
19  &  2019-11-28 & 2019-12-23 & 17060 & 676.1 & 2 & 0   \vspace{0.2cm}  \\ 
20  &  2019-12-24 & 2020-01-20 & 17626 & 678.9 & 0 & 0   \vspace{0.2cm}  \\ 
21  &  2020-01-21 & 2020-02-18 & 18742 & 695.5 & 0 & 0   \vspace{0.2cm}  \\ 
22  &  2020-02-19 & 2020-03-17 & 18607 &  709.6 & 1 & 1   \vspace{0.2cm}  \\ 
23  &  2020-03-19 & 2020-04-15 & 17508 & 710.1 & 1 & 0   \vspace{0.2cm}  \\ 
24  &  2020-04-16 & 2020-05-12 & 18206 & 752.7 & 2 & 0   \vspace{0.2cm}  \\ 
25  &  2020-05-14 & 2020-06-08 & 17233 & 795.3 & 0 & 0   \vspace{0.2cm}  \\ 
26  &  2020-06-09 & 2020-07-04 & 16939 & 805.9 & 0 & 0   \vspace{0.2cm}  \\ 
40  &  2021-06-24 & 2021-07-23 & 19607 & 698.4 & 0 & 0   \vspace{0.2cm}  \\ 
41  &  2021-07-23 & 2021-08-20 & 18321 & 683.5 & 0 & 0   \vspace{0.2cm}  \\ 
47  &  2021-12-30 & 2022-01-28 & 18861 & 685.5 & 0 & 1   \vspace{0.2cm}  \\ 
49  &  2022-02-26 & 2022-03-26 & 17959 & 713.2 & 2 & 2   \vspace{0.2cm}  \\ 
50  &  2022-03-26 & 2022-04-22 & 17186 & 717.1 & 2 & 0   \vspace{0.2cm}  \\ 
51  &  2022-04-22 & 2022-05-18 & 16421 & 738.7 & 2 & 0   \vspace{0.2cm}  \\ 
52  &  2022-05-18 & 2022-06-13 & 16745 & 822.6 & 0 & 0   \vspace{0.2cm}  \\ 
53  &  2022-06-13 & 2022-07-09 & 17304 & 718.5 & 0 & 0   \vspace{0.2cm}  \\ 
54  &  2022-07-09 & 2022-08-05 & 17895 & 712.5 & 0 & 0   \vspace{0.2cm}  \\ 
56  &  2022-09-01 & 2022-09-30 & 19609 & 660.5 & 1 & 1   \vspace{0.2cm}  \\ 
57  &  2022-09-30 & 2022-10-29 & 17990 & 662.6 & 0 & 0   \vspace{0.2cm}  \\ 
59  &  2022-11-26 & 2022-12-23 & 18534 & 692.3 & 2 & 0   \vspace{0.2cm}  \\ 
60  &  2022-12-23 & 2023-01-18 & 14253 & 693.2 & 0 & 0   \vspace{0.2cm}  \\ 
73  &  2023-12-07 & 2024-01-03 & 17296 & 707.0 & 1 & 0   \vspace{0.2cm}  \\ 
74  &  2024-01-03 & 2024-01-30 & 18783 & 711.3 & 1 & 0   \vspace{0.2cm}  \\ 
75  &  2024-01-30 & 2024-02-26 & 19472 & 686.2 & 0 & 0   \vspace{0.2cm} \\ 
76  &  2024-02-26 & 2024-03-26 & 19021 & 697.1 & 0 & 0  \vspace{0.1cm} \\ \bottomrule
\bottomrule
\end{tabular}
\begin{tablenotes}
\item[1] The columns 'Additional transits' indicate the number of transits missed in the PDCSAP data of each sector compared to our custom light curve extraction, both for planets b and c.
\end{tablenotes}
\end{threeparttable}
\end{table*}

\begin{table*}[h!]
\centering
\caption{First five entries of the HARPS-N spectroscopic data.}
\label{tab:RV}
\begin{tabular}{@{}llllllllllp{1cm}@{}}
\toprule
Time   &      RV     & $\sigma_{RV}$ & FWHM & $\sigma_{FWHM}$ & \rhk & $\sigma_{\rhk}$ & BIS & $\sigma_{BIS}$ & $H_{\alpha}$ & $\sigma_{H_{\alpha}}$ \\ 
$[$BJD-2400000$]$ & [m/s] & [m/s] & [m/s] & [m/s] & & & [m/s] & [m/s] & & \\ \midrule 
% jdb   vrad    svrad   fwhm    sig_fwhm        rhk     sig_rhk bis_span        sig_bis_span    ha      sig_ha  ins_name
% ---   ----    -----   ----    --------        --- -------     --------        ------------    --      ------  --------
58920.663142 & 1.856 & 1.114 & 5.7477 & 0.0041 & -4.924 & 0.013 & -0.0085 & 0.0017 & 0.3062 & 0.0011 \\ 
58924.716634 & 5.417 & 1.494 & 5.7480 & 0.0050 & -4.926 & 0.022 & -0.0079 & 0.0021 & 0.3091 & 0.0015 \\ 
58974.555521 & 2.768 & 1.190 & 5.7402 & 0.0034 & -4.944 & 0.013 & -0.0082 & 0.0014 & 0.3091 & 0.0013 \\
58975.562179 & 0.713 & 1.174 & 5.7401 & 0.0035 & -4.952 & 0.015 & -0.0082 & 0.0014 & 0.2983 & 0.0011 \\ 
58976.532753 & 0.210 & 0.898 & 5.7461 & 0.0038 & -4.947 & 0.010 & -0.0066 & 0.0016 & 0.3055 & 0.0011 \\ 
... & & & & & & & & & & \\ \bottomrule
\bottomrule
\end{tabular}
\begin{tablenotes}
\item The full table is available at the CDS. For each observation, we provide the time of the measurement in column 1, the RV estimation together with its uncertainty in columns 2-3, and activity indicators with their uncertainties in columns 4-11 (the CCF FWHM, \rhk, CCF bisector span, and \ha, respectively).
\end{tablenotes}
\end{table*}

\FloatBarrier %\usepackage{placeins}

\section{Interstellar absorption}
\begin{figure*}[h!]
\vspace*{1 cm}
    \centering
\includegraphics[width=\linewidth]{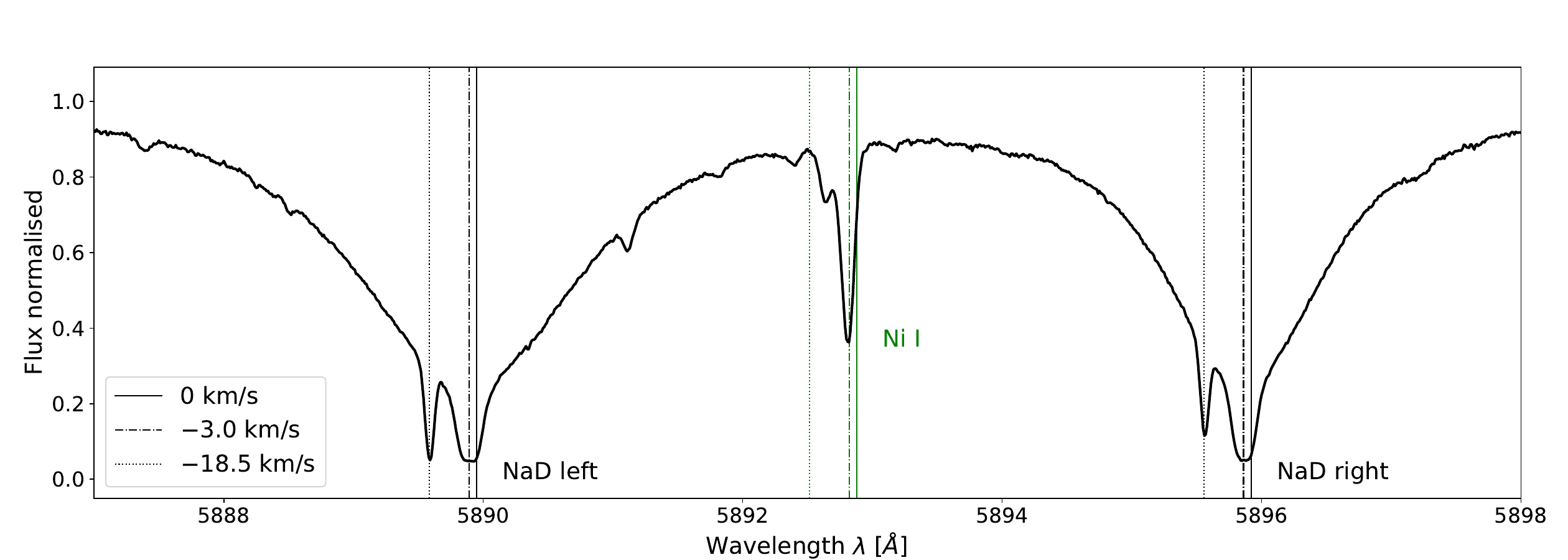}
\caption{HARPS-N spectrum zoomed onto the NaI doublet. The main absorption lines contain an additional narrow feature blueshifted with respect to their centre. These secondary lines are not visible in other spectral lines. Therefore, we exclude contamination from the nearby stellar companion as the cause of this additional absorption. These features have a constant relative position with respect to the main absorption lines of $\sim$15.5 km/s. We attribute this observation to interstellar absorption.} 
\label{Fig:NaI_absorption} 
\end{figure*}

During our analysis of the HARPS-N spectra, we found narrow absorption features in the wide NaD lines around 5890 and 5896 ${\AA}$. In Fig. \ref{Fig:NaI_absorption}, we present the section of a spectrum where this observation was made. This feature is likely the absorption signature of the interstellar medium itself on the Sodium line since the same feature is observed on both Sodium Doublet lines and is for instance not visible in the Ni1 line between the Sodium lines.

Notably, this feature appears  when the observed star is sufficiently far away (here $\sim$80 pc) to exhibit a large extinction \citep[see Fig.3 in][]{Strassmeier2020}. We note that the intensity of such absorption is sky-direction-type dependent since the gas density is not homogeneously distributed \citep[see Fig.2 in][]{Murga2015}. By computing the RV shift of the feature compared to the wavelength in the stellar rest-frame, we estimated that the velocity of the interstellar medium along the line-of-sight was about -15.5 km/s which is compatible with the expected kinematics in the Galactic plane.
The line profile itself is very different from the stellar line profiles, with a thinner profile in agreement with the expected absorption by a cool material containing small micro-turbulence. 

\FloatBarrier %\usepackage{placeins}

\section{Final system parameters}
We report in Table \ref{tab:PlanetParam} the results of our final joint fit of the HARPS-N and TESS data. 

\begin{table*}[h!]
\centering
\begin{threeparttable}
\caption{TOI-1453 system parameters from our joint fit of HARPS-N and TESS observations.}
\label{tab:PlanetParam}
\begin{tabular}{@{}llccccp{6cm}@{}}
\toprule
Parameter  &   Description  &   Prior    &   &  Value    \\ \midrule 
\multicolumn{2}{l}{\textit{\underline{Instrumental parameters}}} \vspace{0.1cm} \\ 
$q_1$  & Limb darkening coef. &  $\mathcal{N}$(0.3911, 0.05) &   & 0.379$\substack{+0.043 \\ -0.047}$  \vspace{0.1cm} \\   
$q_2$ & Limb darkening coef. &   $\mathcal{N}$(0.3358, 0.05)  & & 0.327$\substack{+0.050 \\ -0.047}$  \vspace{0.1cm} \\   
$\delta_{flux}$ & Flux offset & $\mathcal{N}$(0, 0.1) & & -0.0000193(15) \vspace{0.1cm} \\
$D$ & Dilution factor & Fixed & & 0.98435  \vspace{0.1cm} \\ 
$\sigma_{TESS}$ [ppm] & Phot. jitter & $\log\mathcal{U}$(0.1, 1000) & & 1.0$\substack{+3.8 \\ -0.8}$ \vspace{0.1cm} \\ 
% RV parameters
$\sigma_{HARPSN}$ [m s$^{-1}$] & RV jitter & $\mathcal{U}$(0, 10)  & & 2.20$\substack{+0.21 \\ -0.19}$  \vspace{0.1cm} \\  
$\mu_{HARPSN}$ [m s$^{-1}$]  & RV offset & $\mathcal{U}$(-10, 10)  & & 0.07$\pm$0.27  \vspace{0.35cm} \\  
\multicolumn{2}{l}{\textit{\underline{Stellar parameters}}} \vspace{0.1cm} \\ 
$\rho$ [kg m$^{-3}$]  & Stellar density & $\mathcal{N}$(2700, 200) & & 2720.8$\substack{+198.4 \\ -198.8}$  \vspace{0.35cm} \\ 
\multicolumn{2}{l}{\textit{\underline{Planet parameters}}} \vspace{0.1cm} \\ 
$P_b$ [days]    & Orbital period  & $\mathcal{N}$(4.3136, 0.01) & & 4.3135225$\substack{+0.0000118 \\ -0.0000092}$  \vspace{0.07cm} \\
$P_c$ [days]  & & $\mathcal{N}$(6.5887, 0.01) & & 6.5886982(41) \vspace{0.1cm} \\
$T_{C, b}$ [BJD-2459390]  &  Time of conjunction  & $\mathcal{N}$(2.9088,0.01) & & 2.90013$\substack{+0.00083 \\ -0.00093}$ \vspace{0.07cm} \\
$T_{C, c}$ [BJD-2459390] & & $\mathcal{N}$(5.3735, 0.01) & & 5.38133(39) \vspace{0.1cm} \\
$r_{1,b}$ & & $\mathcal{U}$(0, 1) & & 0.572$\substack{+0.038 \\ -0.048}$  \vspace{0.07cm}  \\ 
$r_{2,b}$ & & $\mathcal{U}$(0, 1) & & 0.124$\pm$0.011  \vspace{0.07cm}  \\ 
$r_{1,c}$ & & $\mathcal{U}$(0, 1) & & 0.765$\pm$0.014  \vspace{0.07cm}  \\ 
$r_{2,c}$ & & $\mathcal{U}$(0, 1) & & 0.458$\pm$0.009 \vspace{0.1cm}  \\ 
$e_b$ & Orbital eccentricity & Fixed & & 0 \vspace{0.07cm} \\ 
$e_c$ &  & Fixed & & 0 \vspace{0.1cm} \\ 
$\omega_b$ [deg] & Argument of periastron & Fixed & & 90 \vspace{0.07cm} \\ 
$\omega_c$ [deg] &  & Fixed & & 90 \vspace{0.1cm} \\ 
$K_b$ [m s$^{-1}$] & RV semi-amplitude & $\mathcal{U}$(0, 100) & & 0.61$\pm$0.32 \vspace{0.07cm} \\
$K_c$ [m s$^{-1}$] &  & $\mathcal{U}$(0, 100) & & 1.26$\pm$0.35 \vspace{0.35cm} \\
\multicolumn{2}{l}{\textit{\underline{Derived parameters}}} \vspace{0.1cm} \\ 
$a_b / R_{\star}$ & Scaled semi-major axis & & & 13.88$\substack{+0.33 \\ -0.35}$ \vspace{0.07cm} \\ 
$a_c / R_{\star}$ &  & & & 18.41$\substack{+0.44 \\ -0.46}$ \vspace{0.1cm} \\ 
$R_b / R_{\star}$ & Planet-to-star radius ratio & & & 0.01496$\substack{+0.00041 \\ -0.00042}$ \vspace{0.07cm} \\ 
$R_c / R_{\star}$ &  & & & 0.02832(36) \vspace{0.1cm} \\ 
$b_b$ & Impact parameter & & & 0.569$\substack{+0.039 \\ -0.050}$ \vspace{0.07cm} \\ 
$b_c$ &  & & & 0.767$\pm$0.014 \vspace{0.1cm} \\ 
$i_b$ [deg] & Orbital inclination & & & 87.65$\substack{+0.24 \\ -0.22}$ \vspace{0.07cm} \\ 
$i_c$ [deg] &  & & & 87.61$\substack{+0.10 \\ -0.11}$ \vspace{0.1cm} \\ 
$T_{14, b}$ [hours] & Total transit duration & & & 1.998$\substack{+0.075 \\ -0.068}$ \vspace{0.07cm} \\ 
$T_{14, c}$ [hours] &  & & & 1.874$\substack{+0.108 \\ -0.101}$ \vspace{0.1cm} \\ 
$R_b$ [$R_{\oplus}$] & Planet radius & & & 1.173$\substack{+0.059 \\ -0.056}$ \vspace{0.07cm} \\ 
$R_c$ [$R_{\oplus}$] &  & & & 2.224$\pm$0.095 \vspace{0.1cm} \\ 
$m_b$ [$M_{\oplus}$] & Planet mass & & & 1.24$\substack{+0.66 \\ -0.65}$ \vspace{0.07cm} \\ 
 &  & & & <2.32 (99$\%$) \vspace{0.07cm} \\ 
$m_c$ [$M_{\oplus}$] &  & & & 2.95$\substack{+0.83 \\ -0.84}$ \vspace{0.1cm} \\ 
$\rho_b$ [$\rho_{\oplus}$] & Planet bulk density & & & 0.76$\substack{+0.44 \\ -0.40}$ \vspace{0.07cm} \\ 
 &  & & & <1.49 (99$\%$) \vspace{0.07cm} \\ 
$\rho_c$ [$\rho_{\oplus}$] &  & & & 0.267$\substack{+0.087 \\ -0.080}$ \vspace{0.1cm} \\ \bottomrule
\bottomrule
\end{tabular}
\end{threeparttable}
\end{table*}

\FloatBarrier %\usepackage{placeins}

\section{Internal structure of TOI-1453 c}
Table \ref{tab:int_struct_c} details the posterior resulting from the internal structure modelling of TOI-1453\,c. This table contains the mass fractions of the core, mantle and envelope together with the fraction of individual elements in the core and mantle. We report the results for both water-rich and water-poor formation scenarios, and for all three sets of Si/Mg/Fe priors. 

\begin{table*}[h!]
\renewcommand{\arraystretch}{1.5}
\vspace*{1 cm}
\caption{Results of the internal structure modelling for TOI-1453 c.}
\centering
\begin{tabular}{r|ccc|ccc}
\hline \hline
Water prior &              \multicolumn{3}{c|}{Water-rich (formation outside iceline)} & \multicolumn{3}{c}{Water-poor (formation inside iceline)} \\
Si/Mg/Fe prior &           Stellar (A1) &       Iron-enriched (A2) &      Free (A3) &
                           Stellar (B1) &       Iron-enriched (B2) &      Free (B3) \\
\hline
w$_\textrm{core}$ [\%] &        $18_{-12}^{+14}$ &    $26_{-18}^{+20}$ &    $12_{-9}^{+16}$ &
                           $26_{-18}^{+18}$ &    $37_{-26}^{+27}$ &    $17_{-12}^{+21}$ \\
w$_\textrm{mantle}$ [\%] &      $49_{-14}^{+18}$ &    $42_{-19}^{+21}$ &    $53_{-14}^{+19}$ &
                           $74_{-18}^{+18}$ &    $62_{-27}^{+26}$ &    $83_{-21}^{+12}$ \\
w$_\textrm{envelope}$ [\%] &    $31.9_{-19.6}^{+14.1}$ &    $30.7_{-18.8}^{+15.0}$ &    $32.7_{-20.0}^{+13.5}$ &
                           $0.3_{-0.1}^{+0.2}$ &    $0.6_{-0.2}^{+0.3}$ &    $0.2_{-0.1}^{+0.3}$ \\
\hline
Z$_\textrm{envelope}$ [\%] &        $89.8_{-12.4}^{+5.9}$ &    $87.6_{-14.2}^{+5.8}$ &    $90.9_{-11.6}^{+6.5}$ &
                           $0.5_{-0.2}^{+0.2}$ &    $0.5_{-0.2}^{+0.2}$ &    $0.5_{-0.2}^{+0.2}$ \\
\hline
x$_\textrm{Fe,core}$ [\%] &     $90.4_{-6.4}^{+6.5}$ &    $90.4_{-6.4}^{+6.5}$ &    $90.3_{-6.4}^{+6.5}$ &
                           $90.4_{-6.4}^{+6.5}$ &    $90.4_{-6.4}^{+6.5}$ &    $90.3_{-6.4}^{+6.5}$ \\
x$_\textrm{S,core}$ [\%] &      $9.6_{-6.5}^{+6.4}$ &    $9.6_{-6.5}^{+6.4}$ &    $9.7_{-6.5}^{+6.4}$ &
                           $9.6_{-6.5}^{+6.4}$ &    $9.6_{-6.5}^{+6.4}$ &    $9.7_{-6.5}^{+6.4}$ \\
\hline
x$_\textrm{Si,mantle}$ [\%] &   $34_{-8}^{+11}$ &    $20_{-8}^{+14}$ &    $35_{-24}^{+29}$ &
                           $34_{-8}^{+11}$ &    $20_{-8}^{+14}$ &    $36_{-24}^{+29}$ \\
x$_\textrm{Mg,mantle}$ [\%] &   $34_{-9}^{+11}$ &    $20_{-9}^{+15}$ &    $37_{-25}^{+30}$ &
                           $34_{-9}^{+11}$ &    $21_{-9}^{+15}$ &    $37_{-25}^{+30}$ \\
x$_\textrm{Fe,mantle}$ [\%] &   $32_{-19}^{+13}$ &    $59_{-29}^{+16}$ &    $20_{-15}^{+24}$ &
                           $32_{-19}^{+13}$ &    $59_{-29}^{+15}$ &    $19_{-14}^{+23}$ \\
\hline
\end{tabular}
\label{tab:int_struct_c}
\end{table*}
\renewcommand{\arraystretch}{1.0}

\end{appendix}

\end{document}